\pdfoutput=1
\documentclass[prb,twocolumn]{revtex4}
\usepackage{graphicx,amsmath,amssymb,amsfonts,latexsym,color,dcolumn,bm,mathtools}
\usepackage{verbatim}
\usepackage{epstopdf}

\providecommand{\moy}[1]{\langle #1 \rangle}

\definecolor{green}{rgb}{0,0.498,0}
\definecolor{blue}{rgb}{0.223,0.223,0.667}
\definecolor{red}{rgb}{0.7,0,0}
\definecolor{purp}{rgb}{0.5,0,0.5}
\definecolor{teal}{rgb}{0,0.5,0.5}

\begin{document}

\title{Resolving the vacuum fluctuations of an optomechanical system using an artificial atom}

\author{F. Lecocq, J. D. Teufel, J. Aumentado,  R. W. Simmonds}

\affiliation{National Institute of Standards and Technology, 325 Broadway, Boulder, CO 80305, USA}

\date{\today}

\maketitle

\textbf{Heisenberg's uncertainty principle results in one of the strangest quantum behaviors: an oscillator can never truly be at rest. Even in its lowest energy state, at a temperature of absolute zero, its position and momentum are still subject to quantum fluctuations \cite{Clerk2010,Aspelmeyer2013}. Resolving these fluctuations using linear position measurements is complicated by the fact that classical noise can masquerade as quantum noise \cite{Jayich2012,Safavi-Naeini2012,Safavi-Naeini2013,Weinstein2014}. On the other hand, direct energy detection of the oscillator in its ground state makes it appear motionless \cite{O'Connell2010,Clerk2010}. So how can we resolve quantum fluctuations? Here, we parametrically couple a micromechanical oscillator to a microwave cavity to prepare the system in its quantum ground state \cite{Teufel2011b,Palomaki2013} and then amplify the remaining vacuum fluctuations into real energy quanta \cite{Palomaki2013a}. Exploiting a superconducting qubit as an artificial atom, we measure the photon/phonon-number distributions \cite{Brune1996,Meekhof1996,Hofheinz2008} during these optomechanical interactions. This provides an essential non-linear resource to, first, verify the ground state preparation and second, reveal the quantum vacuum fluctuations of the macroscopic oscillator's motion. Our results further demonstrate the ability to control a long-lived mechanical oscillator using a non-Gaussian resource, directly enabling applications in quantum information processing and enhanced detection of displacement and forces.
}

Cavity optomechanical systems have emerged as an ideal testbed for exploring the quantum limits of linear measurement of macroscopic motion\cite{Aspelmeyer2013}, as well as a promising new architecture for performing quantum computations. In such systems, a light field reflecting off a mechanical oscillator acquires a position-dependent phase shift and reciprocally, it applies a force onto the mechanical oscillator. This effect is enhanced by embedding the oscillator inside a high-quality factor electromagnetic cavity. Numerous physical implementations exist, both in the microwave and optical domain, and have been used to push the manipulation of macroscopic oscillators into the quantum regime, demonstrating laser cooling to the ground state of motion\cite{Teufel2011b, Chan2011}, coherent transfer of itinerant light fields into mechanical motion \cite{Verhagen2012,Palomaki2013}, or their entanglement\cite{Palomaki2013a}. Thus far, linear position measurements have provided evidence for the quantization of light fields via radiation pressure shot noise\cite{Purdy2013b,Suh2014} and mechanical vacuum fluctuations via motional sideband asymmetries \cite{Jayich2012,Safavi-Naeini2012,Safavi-Naeini2013,Weinstein2014}. However, the use of only classical and linear tools has restricted most optomechanical experiments to the manipulation of Gaussian states.

The addition of a strong non-linearity, such as an atom, has fostered tremendous progress towards exquisite control over non-Gaussian quantum states of light fields and atomic motion\cite{Brune1996,Meekhof1996}. First developed in the context of cavity quantum electrodynamics, these techniques are now widely applied to engineered systems, such as superconducting quantum bits (qubits) and microwave resonant circuits\cite{Wallraff2004,Sillanpaa2007,Hofheinz2008,Hofheinz2009,Pirkkalainen2013}. In a pioneering experiment incorporating a high-frequency mechanical oscillator\cite{O'Connell2010}, single phonon Fock-state control was demonstrated, although short energy life-times of the mechanical oscillator and the qubit have slowed any further progress.

 In this work, we develop a unique architecture that incorporates an artificial atom --a superconducting qubit\cite{Martinis2009a}-- into a circuit cavity electromechanical system\cite{Teufel2011}, on a single chip. Here, a low-frequency, high quality factor mechanical oscillator strongly interacts with the microwave cavity photons. The qubit-cavity interaction realizes a non-classical emitter and detector of photons, thus providing an essential non-linear resource for the deterministic control of long-lived mechanical quantum states. We demonstrate the potential of such an architecture by measuring the quantum vacuum fluctuations inherently present in the motion of a macroscopic oscillator.

\begin{figure*}
\includegraphics[scale=1]{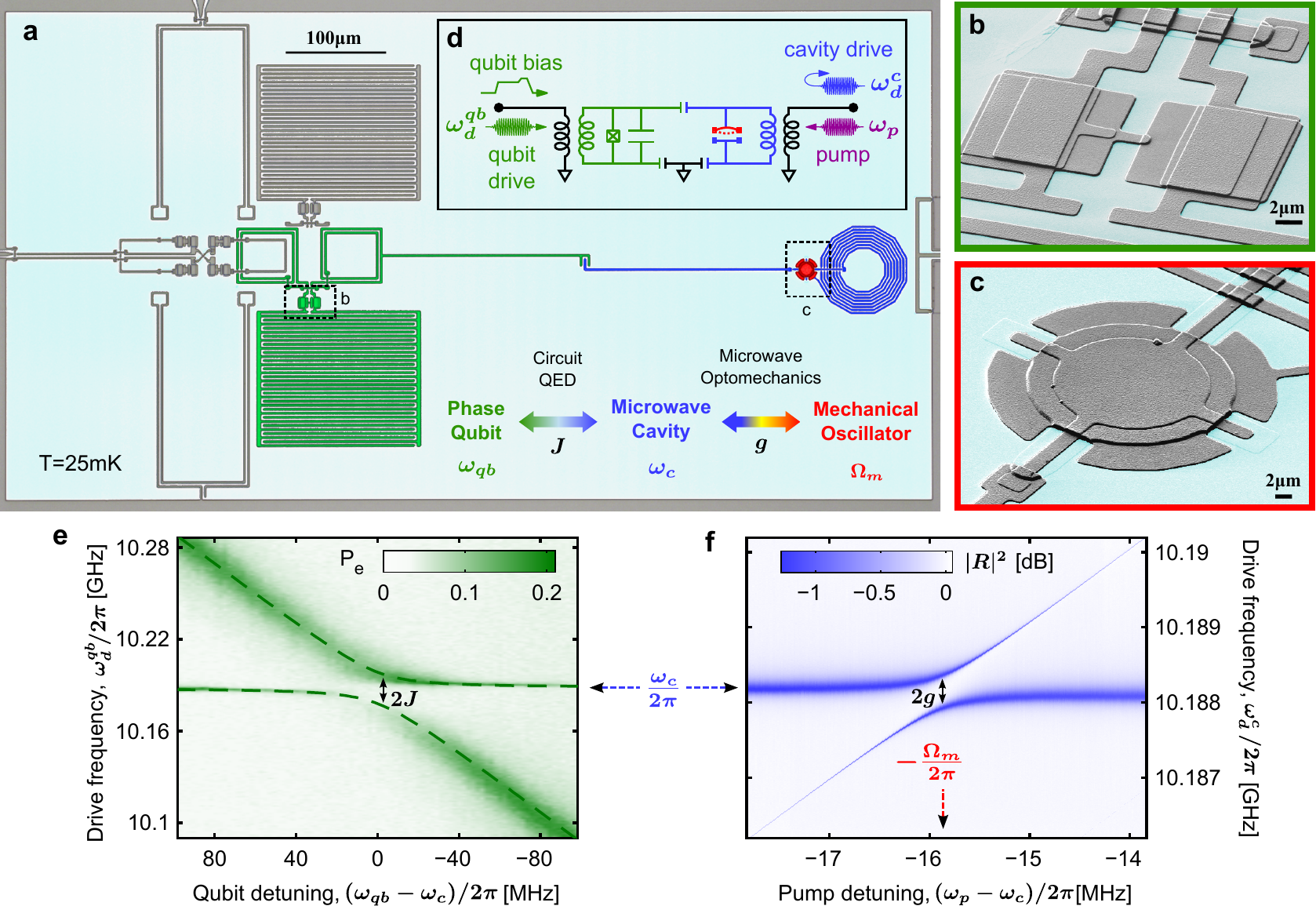}
\caption{\textbf{Device description and strong coupling regime}. \textbf{a}, False-color optical micrograph of the device. Aluminum is in grey and sapphire in light blue. The phase qubit is in green; the microwave cavity in blue; and the mechanically compliant capacitor in red. \textbf{b,c}, False-color scanning electron micrograph of the qubit's Josephson junction and of the mechanical oscillator, respectively. \textbf{d}, Equivalent circuit diagram. \textbf{e}, Qubit/cavity vacuum Rabi splitting. Population of the qubit's excited state $P_{e}$, in green, as a function of the drive frequency and of the qubit detuning with respect to the microwave cavity. The avoided crossing is a clear signature of the single-photon strong coupling regime. From a fit with theory we extract the qubit/cavity coupling rate $J/2\pi=12.5~\rm{MHz}$. \textbf{f}, Cavity/oscillator normal-mode splitting. Reflected power in blue as a function of the drive and pump frequencies. The drive is applied near the cavity resonance while the pump is applied near $\Delta_p=-\Omega_m$. Here, the pump strength is set to $n_p\approx5\times10^5$. The normal-mode splitting observed in the cavity driven response corresponds to the hybridization of the cavity and mechanical modes in the driven strong coupling regime. From a fit with theory we extract $\kappa/2\pi=163~\rm{kHz}$, $\Gamma_{m}/2\pi=150~\rm{Hz}$ and $g=2\pi\times242~\rm{kHz}>(\bar{n}_m\Gamma_{m},\kappa)$ where $\bar{n}_m\approx32$ at $T\approx25~\rm{mK}$.
\label{fig1}}
\end{figure*}

A microwave cavity is the central element of this architecture (in blue in Fig.\ref{fig1}a). It is a linear inductor-capacitor (LC) resonator formed by a coil inductor and a mechanically compliant vacuum-gap capacitor \cite{Cicak2010,Teufel2011}. First, the intra-cavity electromagnetic field is coupled via radiation pressure to the vibrational mode of the compliant capacitor (in red in Fig.\ref{fig1}). Second, the microwave cavity is capacitively coupled to a phase qubit (in green in Fig.\ref{fig1}). A  phase qubit is formed from a Josephson junction in parallel with an LC oscillator, and it behaves like a non-linear resonator at the single quantum level, \textit{i.e}, an artificial atom \cite{Martinis2009a}. To a good approximation the phase qubit can be operated as a two-level system  whose transition frequency $\omega_{qb}$ can be widely tuned \textit{in situ} by applying an external flux, such that $9~\rm{GHz} \le \omega_{qb}/2\pi \le 13.5~\rm{GHz}$. The microwave cavity and the fundamental flexural mode of the capacitor are two harmonic oscillators with resonance frequencies of respectively $\omega_{c}/2\pi= 10.188~\rm{GHz}$ and $\Omega_{m}/2\pi= 15.9~\rm{MHz}$.

The qubit and the cavity are both electrical circuits with quantized energy levels, sharing a voltage through the coupling capacitor. On resonance, $\Delta_{qb}=\omega_{qb}-\omega_{c}=0$, the interaction between the qubit and the cavity is well described by the Jaynes-Cummings Hamiltonian $\mathcal{H}_{jc}= \hbar J\left(\hat{a}\hat{\sigma}_{+}+ \hat{a}^{\dagger} \hat{\sigma}_{-}\right)$. Here, $\hat{\sigma}_{+}$ ($\hat{\sigma}_{-}$) is the raising (lowering) operator for the qubit, $\hat{a}^{\dagger}$ ($\hat{a}$) is the creation (annihilation) operator for cavity photons and $J$ is the capacitive coupling strength. $\mathcal{H}_{jc}$ describes the exchange of a single quantum between the qubit and the cavity at a rate $2J$. In the strong coupling regime, when the coupling strength $J$ overcomes the decoherence rates of the qubit $\gamma_{qb}$ and the cavity $\kappa$, \textit{i.e} $J>(\gamma_{qb},\kappa)$, the system hybridizes, leading to the well known vacuum Rabi splitting, measured spectroscopically in Fig.\ref{fig1}e. The qubit-cavity interaction can be effectively turned off by detuning the qubit.

The position of the mechanical oscillator modulates the cavity resonance frequency and thus, the energy stored in the cavity. As a result, the microwave photons apply a force on the mechanical oscillator. This interaction is described by the \textit{radiation pressure} Hamiltonian \cite{Aspelmeyer2013} $\mathcal{H}_{rp}=\hbar G\hat{n}_{c}\hat{x}$, where $G=d \omega_{c}/dx$, $ \hat{n}_{c}= \hat{a}^{\dagger} \hat{a}$ is the cavity photon number and $\hat{x} = x_{zpf}\left(\hat{b}^{\dagger}+\hat{b}\right)$ is the oscillator's position. Here, $x_{zpf}$ is the oscillator's zero point fluctuation and $\hat{b}^{\dagger}$ ($\hat{b}$) is the creation (annihilation) operator for mechanical phonons. The force applied by a single photon onto the mechanical oscillator is typically weak, with $g_0= Gx_{zpf}\ll (\bar{n}_m\Gamma_{m},\kappa)$ where $\bar{n}_m$ is the equilibrium thermal occupancy of the oscillator and $\Gamma_{m}$ its intrinsic relaxation rate. However the total force increases significantly with the intensity of the intra-cavity field. In the presence of a strong coherent microwave pump of frequency $\omega_p$, the optomechanical interaction is linearized and takes two different forms depending on the pump-cavity detuning $\Delta_p = \omega_p - \omega_c$ (Methods). When $\Delta_p=-\Omega_m$, the annihilation of a mechanical phonon can up-convert a pump photon into a cavity photon, mediating a ``beam splitter'' interaction, $\mathcal{H}_{-}=\hbar g\left( \hat{a} \hat{b}^{\dagger} + \hat{b} \hat{a}^{\dagger}\right)$. This results in the coherent exchange of the cavity and mechanical states at a rate $2g$, where $g=g_0 \sqrt{n_p}$ is the enhanced optomechanical coupling and $n_p$ is the pump strength expressed in terms of the average number of intra-cavity photons. When $\Delta_p=+\Omega_m$, pump photons are down-converted into correlated photon-phonon pairs,  mediating a ``two-mode squeezer'' interaction, $\mathcal{H}_{+}=\hbar g\left( \hat{a}^{\dagger} \hat{b}^{\dagger} +  \hat{b} \hat{a} \right)$. This results in the amplification and entanglement of the cavity field and the mechanical motion\cite{Palomaki2013a}, at a rate $2g$. The hallmark for entering the strong coupling regime in our device, $g>(\Gamma_{m},\kappa)$, is the hybridization and normal-mode splitting induced by a strong beam splitter interaction\cite{Teufel2011}, as measured through the cavity driven response in Fig.\ref{fig1}f. Finally with a lifetime of the mechanical oscillator's ground state, $1/\bar{n}_m\Gamma_m\approx33~\rm{\mu s}$, much longer than the cavity lifetime, $1/\kappa\approx1~\rm{\mu s}$, this device is in the quantum coherent regime\cite{Verhagen2012}.
  
Measurements in the frequency domain provide an extensive characterization of the device's parameters, however, they only probe the steady state of the system, in equilibrium with the thermal environment. In the next two paragraphs, we will show that time domain protocols enable: (1) the preparation of non-classical cavity states and the measurement of the intra-cavity photon-number distribution\cite{Brune1996,Meekhof1996,Hofheinz2008}, and (2) coherent state transfer by frequency conversion and entanglement by parametric amplification between the microwave cavity and the mechanical oscillator\cite{Zakka-Bajjani2011,Hofer2011}.

\begin{figure}
 \includegraphics[scale=0.91]{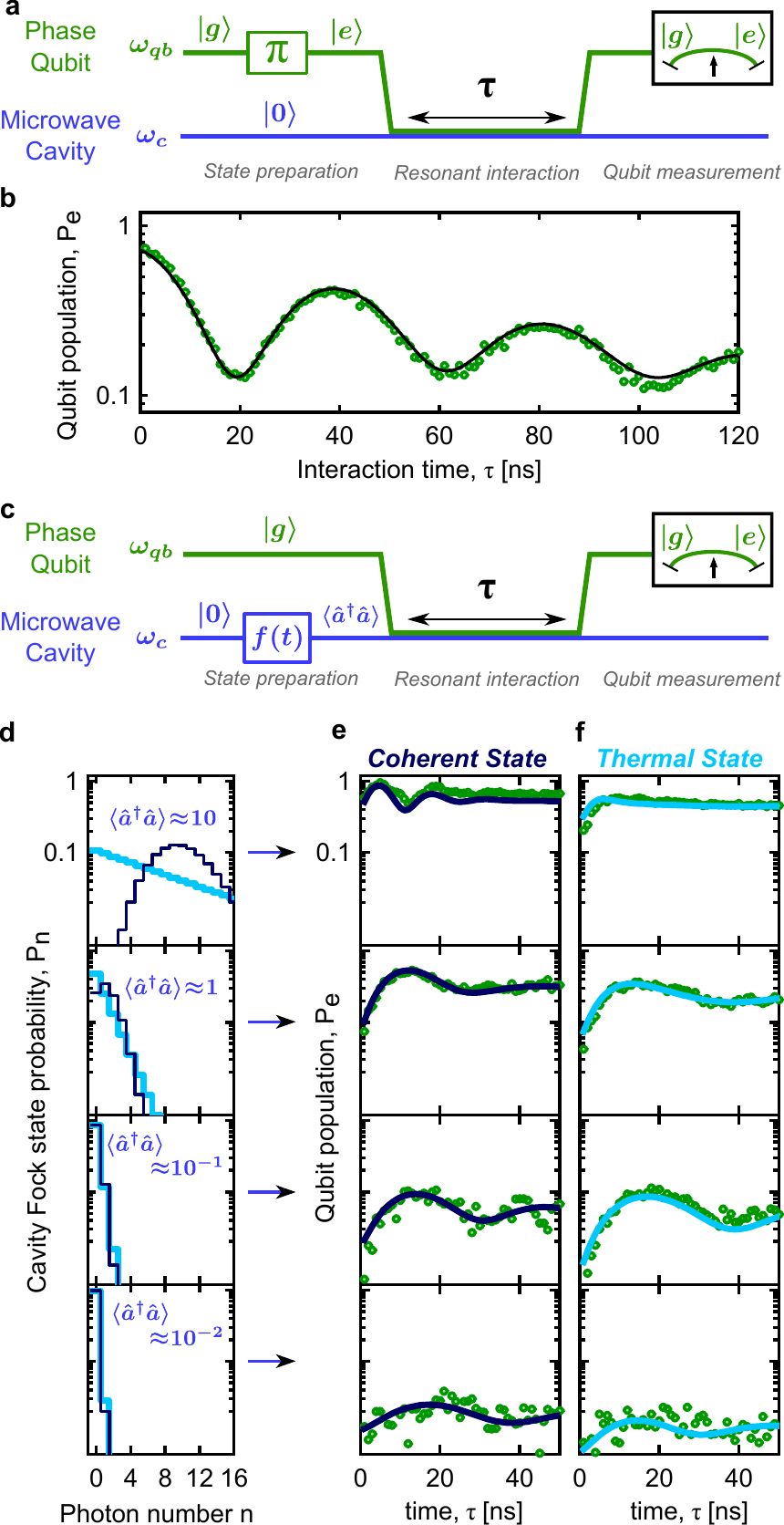}
 \caption{\textbf{Cavity state preparation and readout}. \textbf{a}, Sequence diagram for the preparation of a single photon Fock state in the cavity. The qubit is prepared in the first excited state $|e\rangle$, with a 75\%  efficiency, then interacts resonantly with the cavity for a time $\tau$ before the qubit state is measured. The population of the qubit $P_{e}$ is plotted in \textbf{b} as a function of the interaction time $\tau$. The black line is a fit to a master equation prediction (Methods). \textbf{c}, Sequence diagram for the readout of the cavity state. The cavity is prepared in a coherent (or thermal) state by driving it with a coherent tone (or with white noise). The corresponding intra-cavity photon distributions are shown in \textbf{d} for four drive amplitudes, parametrized by the average cavity occupancy $\bar{n}_c$. The thermal states distributions and coherent states distributions are respectively in bright and dark blue. For each drive amplitude, the evolution of the qubit population $P_{e}$ is plotted in green in \textbf{e} for the coherent states, and in \textbf{f} for the thermal states. The solid lines are a fit to a master equation prediction.
 \label{fig2}}
\end{figure}
  
The out-of-equilibrium dynamics between the phase qubit and the cavity are shown in Fig.\ref{fig2}. First, in Fig.\ref{fig2}a-b we perform the first basic block of the Law and Eberly protocol\cite{Law1996}. We initialize the qubit in the excited state using a resonant microwave pulse, then tune the qubit into resonance with the cavity for an interaction time $\tau$ and measure the qubit population $P_{e}$ (using a destructive single shot readout). The coupled system undergoes vacuum Rabi oscillations at a single frequency $J/\pi$ and after half a cycle the cavity is prepared in a single photon Fock state. Next, in Fig.\ref{fig2}c-f, we exploit the well  known scaling of the Rabi frequency with the cavity Fock state number to measure the intra-cavity photon-number distribution\cite{Hofheinz2008}. We initialize the cavity in either a coherent state or a thermal state, parametrized by the average photon occupancy $\moy{\hat{a}^{\dagger} \hat{a}}$. When the qubit is tuned into resonance, each initial distribution (Fig.\ref{fig2}d) produces a distinct time evolution of $P_e(\tau)$(Fig.\ref{fig2}e-f), in good agreement with simulations that includes all sources of decoherence and where the average photon number $\bar{n}_c$ is the only free parameter (Methods). We resolve the cavity occupancy down to $\moy{\hat{a}^{\dagger} \hat{a}}\approx0.02$ and have the ability to distinguish the thermal, noise-like component of the cavity state from the coherent component.

\begin{figure}
\includegraphics[scale=0.90]{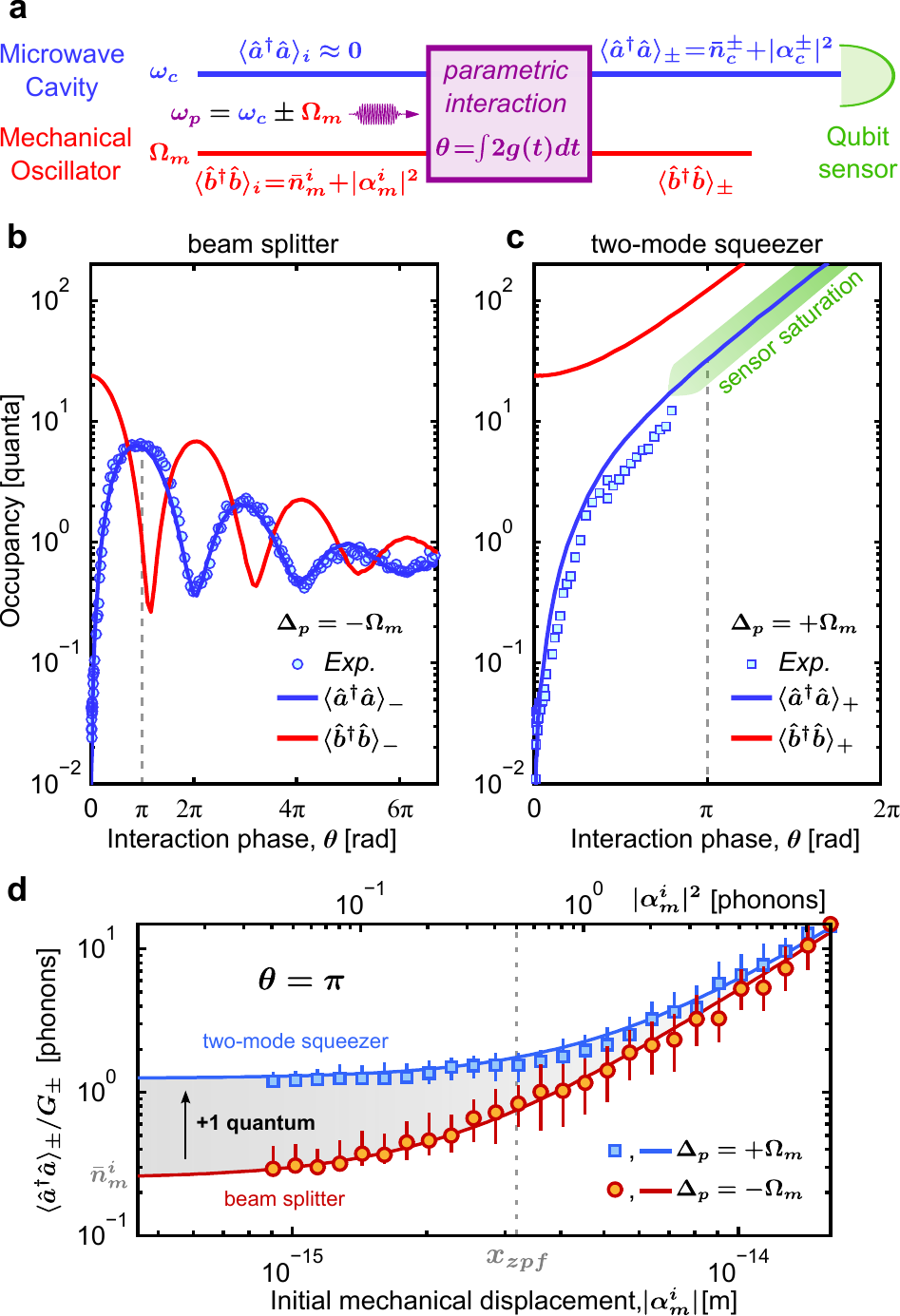}
\caption{\textbf{Optomechanics with a number resolving detector}. \textbf{a}, Sequence diagram. \textbf{b}, Measured cavity occupancy (blue circles) as a function of the interaction duration in reduced unit $\theta$, for $\Delta_p=-\Omega_m$ ($n_p=3.8\times 10^5$ and $g=2\pi\times198~\rm{kHz}$). \textbf{c}, Same as \textbf{b} for $\Delta_p=+\Omega_m$. In both \textbf{b} and \textbf{c} the predictions of the Heisenberg-Langevin equations of motion are shown in solid blue (Methods). The mechanical occupancy is shown as solid red lines. \textbf{d}, Measurement of the vacuum fluctuations of the mechanical oscillator. The cavity occupancy is measured as a function of the initial mechanical displacement for $\theta=\pi$ and for $\Delta_p=\pm\Omega_m$. We define the gain at each pump frequency, $G_{\pm}$, as the ratio of final cavity displacement to initial mechanical displacement $G_{\pm}=|\alpha_c^{\pm}|^2/|\alpha_m^{i}|^2$. We display $\moy{\hat{a}^{\dagger}\hat{a}}_{-}/G_{-}$ in red and $\moy{\hat{a}^{\dagger}\hat{a}}_{+}/G_{+}$ in blue. The amplification of the mechanical vacuum fluctuations appears as one additional quantum when  $\Delta_p=+\Omega_m$.
\label{fig3}}
\end{figure}

We will now exploit this measurement technique to explore the out-of-equilibrium optomechanical dynamics (Fig.\ref{fig3}). To acquire some physical intuition one can solve the lossless equations of motion describing the time evolution of the microwave and mechanical field amplitudes (Methods). The average photon occupancy after a beam splitter interaction, $\moy{\hat{a}^{\dagger}\hat{a}}_{-}$, or a two-mode squeezer interaction,$\moy{\hat{a}^{\dagger}\hat{a}}_{+}$, follow:
 
 \begin{flalign}
 & \moy{\hat{a}^{\dagger}\hat{a}}_{-} =\moy{\hat{a}^{\dagger}\hat{a}}_i \,  \cos^2(\theta/2) \ + \moy{\hat{b}^{\dagger}\hat{b}}_i \, \sin^2(\theta/2)\label{Nred} \\
 & \moy{\hat{a}^{\dagger}\hat{a}}_{+} =\moy{\hat{a}^{\dagger}\hat{a}}_i \cosh^2(\theta/2) + \moy{\hat{b}\hat{b}^{\dagger}}_i\sinh^2(\theta/2) \label{Nblue}
 \end{flalign}
 
 where $\moy{\hat{a}^{\dagger}\hat{a}}_i$ and $\moy{\hat{b}^{\dagger}\hat{b}}_i$ are respectively the initial cavity and mechanical occupancy, and $\theta=\int2g(t)dt$ is the accumulated interaction phase. The periodic functions in Eq.\ref{Nred} describe the state exchange induced by the beam splitter interaction (see Fig.\ref{fig3}b) while the hyperbolic functions in Eq.\ref{Nblue} describe the amplification induced by the two-mode squeezer interaction (see Fig.\ref{fig3}c).
 Experimentally, we start by actively preparing the mechanical state in a nearly pure coherent state, $\moy{\hat{b}^{\dagger}\hat{b}}_i=\bar{n}_m^{i}+|\alpha_m^{i}|^2$,  where $|\alpha_m^{i}|^2=23$ is the coherent component (displacement) and $\bar{n}_m^{i}=0.25$ represents the residual thermal (incoherent) phonon occupancy (Methods). We then pulse either optomechanical interaction using a microwave pump at $\Delta_p =\pm\Omega_m$, followed by tuning the qubit into resonance with the cavity to measure the subsequent photon number distribution (as described previously in Fig.\ref{fig2}) for each pump duration. As expected this distribution corresponds to a displaced thermal state, characterized by an incoherent component $\bar{n}_c^{\pm}$ and a coherent component $\alpha_c^{\pm}$, for a total average photon occupancy of $\moy{\hat{a}^{\dagger}\hat{a}}_{\pm}=\bar{n}_c^{\pm}+|\alpha_c^{\pm}|^2$. In Fig.\ref{fig3}b and c, we display $\moy{\hat{a}^{\dagger}\hat{a}}_{\pm}$ as a function of the interaction phase $\theta$. The data in Fig.\ref{fig3}b (Fig.\ref{fig3}c) qualitatively agree with Eq.\ref{Nred} (Eq.\ref{Nblue}), and quantitatively agree with full numerical simulations (solid blue lines) that include the finite linewidth and bath temperature of each mode (Methods). The expected evolution of $\moy{\hat{b}^{\dagger}\hat{b}}_{\pm}$ follows the solid red line. The only free parameter is the initial mechanical occupancy $\moy{\hat{b}^{\dagger}\hat{b}}_i$. We emphasize our ability to resolve, with a sensitivity well below the single quantum level, the coherent exchange of mechanical phonons and cavity photons or the amplification of the two localized modes, with both processes occurring at a rate faster than decoherence.
 
 A striking signature of the quantum nature of the oscillator's motion resides in the commutation relation $\hat{b}\hat{b}^{\dagger}= \hat{b}^{\dagger}\hat{b}+1$. Together with Eq.\ref{Nblue}, we can see that even with both modes initially in their ground state,$\moy{\hat{a}^{\dagger}\hat{a}}_i=\moy{\hat{b}^{\dagger}\hat{b}}_i=0$, the zero-point motion of the oscillator alone feeds the parametric amplification process, with a gain $\sinh^2(\theta/2)$, leading to a finite cavity occupancy $\moy{\hat{a}^{\dagger}\hat{a}}_{+}$. Whereas, from the same initial conditions, Eq.\ref{Nred} shows no interesting dynamics for the beam splitter interaction, with $\moy{\hat{a}^{\dagger}\hat{a}}_{-}=0$. Thus, to quantitatively extract the ``+1'' contribution during amplification, we compare the two processes, as shown in Fig.\ref{fig3}d. With the cavity initially in its ground state,  $\moy{\hat{a}^{\dagger}\hat{a}}_i=0$, Eq.\ref{Nred} and Eq.\ref{Nblue} relate the final average cavity occupancy $\moy{\hat{a}^{\dagger}\hat{a}}_{\pm}$ to the initial average mechanical occupancy $\moy{\hat{b}^{\dagger}\hat{b}}_i$ through the gain of the parametric interactions, $\sin^2(\theta/2)$ for beam splitter or $\sinh^2(\theta/2)$ for the two-mode squeezer. First, we set the interaction phase to $\theta = \pi$ and measure the final photon distribution as a function of the initial mechanical displacement $\alpha_m^i$, for $\Delta_p =+\Omega_m$ and $\Delta_p =-\Omega_m$. In the presence of loss, the gains are less than their maximum values ($G_-=0.25<\sin^2(\pi/2)$ and $G_+=0.88<\sinh^2(\pi/2)$), but can be measured using large coherent drives, taking the ratio of final cavity displacement to initial mechanical displacement, $G_\pm = |\alpha_c^\pm|^2/|\alpha_m^i|^2$ (Methods). We show the results for $\moy{\hat{a}^{\dagger}\hat{a}}_{-}/G_- = \moy{\hat{b}^{\dagger}\hat{b}}_i$ in red and $\moy{\hat{a}^{\dagger}\hat{a}}_{+}/G_+= \moy{\hat{b}^{\dagger}\hat{b}}_i +1$ in blue. The difference between these two optomechanical interactions is clear in Fig.\ref{fig3}d, showing the extra ``+1'' contribution sourced directly from the commutator between the position and momentum of the macroscopic mechanical oscillator due to the quantum vacuum fluctuations.
  
 This signature has been discussed in terms of asymmetry between the rates of phonon absorption and emission\cite{Safavi-Naeini2012,Weinstein2014}, or in terms of added noise in the context of three-wave mixing\cite{Clerk2010,Bergeal2010a}. Our architecture is however uniquely suited to explore quantitatively such phenomenon in optomechanics, because we have the ability to measure directly the mechanically scattered photons, localized in the cavity. By combining the measurement of the intra-cavity photon-number distribution with the optomechanical interactions, we have realized a \textit{phonon}-number distribution measurement. The non-linearity of the qubit-cavity interaction allows us to clearly distinguish classical noise from quantum noise, as only classical noise can lead to real cavity quanta that can excite the qubit out of its ground state. In addition, we are not sensitive to the correlations between the electromagnetic noise and mechanical noise, which would induce ``squashing'' of the output field \cite{Weinstein2014}.
  
To conclude, by measuring the vacuum fluctuations of an optomechanical system using an artificial atom we have implemented all the tools necessary for the manipulation of quantum states in engineered mechanical systems. We simultaneously achieve long mechanical lifetime, ground-state cooling, and quantum coherent coupling to a non-classical resource. Looking forward, with more complex protocols, we could: (1) exploit the qubit as a deterministic single phonon source to generate arbitrary quantum states of motion \cite{Hofheinz2009}, (2) perform full state tomography of the mechanical oscillator \cite{Leibfried1996} and (3) manipulate photon-phonon entanglement via sequenced beam splitter and two-mode squeezer interactions. The ability to encode complex quantum states in these long-lived mechanical systems has important implications for quantum information and for the study of the fundamental quantum behavior of massive objects\cite{Pikovski2012}.

We thank A. W. Sanders for taking the micrographs in Fig.\ref{fig1}b,c. This work was supported by the NIST Quantum Information Program. Contribution of the U.S. government, not subject to copyright.

\bibliography{Bibliographie_HybridOptoMecha_V16}

\end{document}


\title{Methods for "Resolving the vacuum fluctuations of an optomechanical system using an artificial atom"}

\maketitle

\textbf{Device parameters}
A list of the important parameters and notations is presented in the  Table \ref{DeviceParam}.

\textbf{Qubit operation and Circuit QED theory}
The superconducting qubit used in this experiment is a phase qubit, with a dc-SQUID readout. The concept, operation and limitations are discussed in numerous references \cite{MartinisDevoret2004,Cooper2004,Martinis2009a}. To a good approximation it is a two-level system whose transition frequency $\omega_{qb}$ can be widely tuned in situ by applying an external flux-bias. Its quantum state can be fully controlled using resonant a microwave drive. The state of the qubit is measured destructively, in a single shot, by applying a fast flux-bias pulse that induces the tunneling of the excited state into an adjacent potential well. That well corresponds to a different circulating current inside the qubit inductor and is read-out using a dc-SQUID magnetometer.

\medskip

We develop now the theoretical framework for the qubit/cavity interaction, leading to the use of the qubit as a single photon source and a photon-number distribution detector.
The Hamiltonian describing the coupled system qubit/cavity alone is:

\begin{equation}
\mathcal{H}_{qed}(t)= \frac{1}{2}\hbar\omega_{qb}(t)\hat{\sigma}_Z +\hbar\omega_{c}\left(\hat{a}^{\dagger}\hat{a}+\frac{1}{2}\right) + \hbar J\left(\hat{a}\hat{\sigma}_{+} +\hat{a}^{\dagger}\hat{\sigma}_{-}\right)
\end{equation}

The out-of-equilibrium quantum dynamics in the presence of relaxation and decoherence is captured by the following Lindblad master equation \cite{Haroche2006}:

\begin{equation}
\dot{\rho} (t)=-\frac{i}{\hbar}\left[\mathcal{H}_{qed}(t),\rho(t)\right] 
+\frac{1}{2}\sum_{n} \left[\mathcal{C}_n,\rho(t)\mathcal{C}_n^{\dagger}\right]+ \left[\mathcal{C}_n\rho(t),\mathcal{C}_n^{\dagger}\right]
\label{Lindblad}
\end{equation}

where $\mathcal{C}_n$ are operators describing: qubit relaxation $\mathcal{C}_1 = \sqrt{1/T_1}\hat{\sigma}^{-}$; cavity relaxation $\mathcal{C}_2 = \sqrt{1/T_{1,cav}}\hat{a}$; and qubit pure dephasing $\mathcal{C}_3 = \sqrt{2/T_{\phi}}\hat{\sigma}_{z}$. 

\medskip

\begin{figure*}[t]
\includegraphics[scale=1]{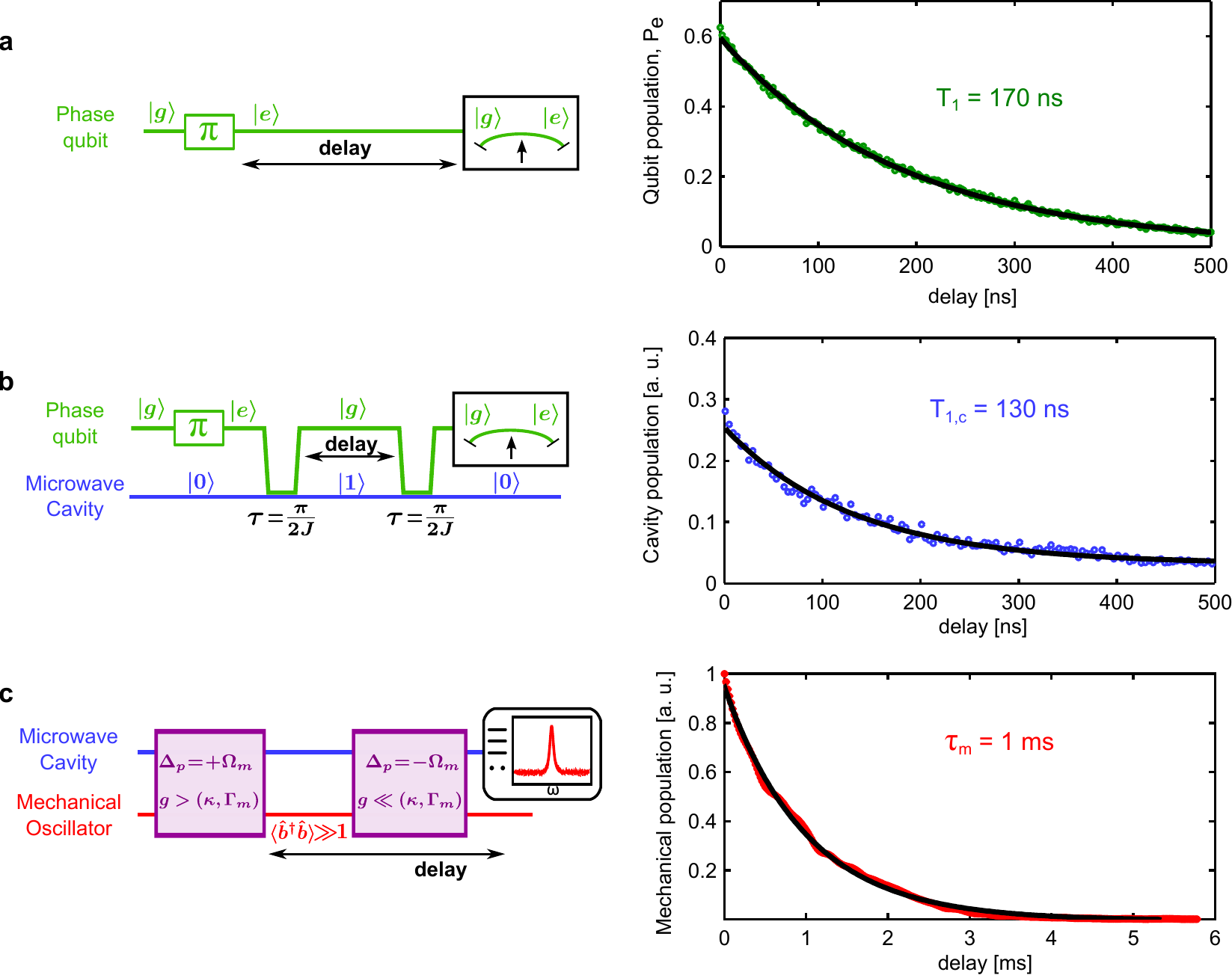}
\caption{\textbf{Phase qubit, microwave cavity and mechanical oscillator relaxation times}. \textbf{a}, Qubit relaxation time. \textbf{b}, Single photon cavity relaxation time measured by storage/retrieval of a single photon Fock state in the cavity. \textbf{c}, Mechanical ``ring-down'' time measured by monitoring the decay of the amplitude of the upper sideband of a weak pump tone at $\Delta_p=\omega_p-\omega_c=-\Omega_m$, for an initially strongly driven mechanical oscillator.
\label{SystemsLifetimes}}
\end{figure*}

To fit the single photon vacuum Rabi oscillation shown in Fig.2a, we numerically integrate Eq.\ref{Lindblad} (using QuTiP \cite{Johansson2013}). We introduce the full time dependence of $\omega_{qb}(t)$ into $\mathcal{H}_{qed}(t)$, including the $4~\rm{ns}$ risetime of the pulse that tunes the qubit into resonance with the cavity, as well as a small linear drift of the qubit frequency during the interaction. We also account for excited state preparation fidelity of about $75\%$ and a measurement contrast of  $51\%$, which agrees with the total visibility of  $~35\%$ measured with the qubit relaxation measurement of Fig.\ref{SystemsLifetimes}a. We use $T_1=160~\rm{ns}$, $T_{1,cav}=110~\rm{ns}$ and $T_{\phi}=45~\rm{ns}$, in good agreement with the independent measurements of relaxation time at the single quantum level, shown in Fig.\ref{SystemsLifetimes}a,b.

\medskip

The fit of the single photon vacuum Rabi oscillations also provides a full calibration of the qubit sensor. The fitting procedure of the qubit evolution used in Fig.2b and Fig.3 can now be strongly constrained to only two parameters: the cavity photon distribution and an eventual qubit flux offset (due to instrument drift and helium level variation) that impacts both the qubit/cavity detuning and the contrast of the qubit state measurement. In Fig.2b the distribution is chosen to be either Poissonian or thermal and the average photon number $\moy{\hat{a}^{\dagger} \hat{a}}$ is the fit parameter. In Fig.3 we more generally consider a displaced thermal state, $\moy{\hat{a}^{\dagger}\hat{a}}=\bar{n}_c+|\alpha_c|^2$, with an incoherent component $\bar{n}_c$ and a coherent component $\alpha_c$. Note that below $\moy{\hat{a}^{\dagger} \hat{a}}\approx10^{-1}$, the Poissonian and thermal distributions are too similar to be distinguished. Also, for $\moy{\hat{a}^{\dagger} \hat{a}}\ge10$, multilevel dynamics of the phase qubit start to induce deviations from the simple two-level model.

\textbf{Optomechanics in the frequency domain}
The microwave cavity's driven response is dressed by its interaction with the mechanical oscillator and gives access to most of the parameters of the system \cite{Teufel2011}. The reflected signal off of the cavity in the presence of a microwave pump, as presented in Fig.1F, follows \cite{Agarwal2010,Weis2010}:

\begin{equation}
  R =  \frac{1}{1+i\alpha_f}-\frac{2\eta\kappa\left(1-i\chi\right)}{\kappa+2i\left(\omega-\omega_c\right)+4\chi\left(\omega_p-\omega_c\right)}
  \label{CavityS11}
\end{equation}
with
\begin{equation}
  \chi = \frac{4g^2\Omega_m}{\left(\kappa+2i(\omega-2\omega_p+\omega_c)\right)\left(\Omega_m^2-(\omega-\omega_p)^2+i(\omega-\omega_p)\Gamma_m\right)}\\
\end{equation}

In the above $\omega_c$ is the microwave cavity frequency; $\Omega_m$ is the mechanical oscillator frequency; $\omega_p$ is the pump frequency; $\Gamma_m$ is the intrinsic mechanical linewidth; $\kappa = \kappa_{int} + \kappa_{ext}$ is the total cavity linewidth, where $\kappa_{ext}$ is the decay rate to the feed line and $\kappa_{int}$ is the decay rate to the environment; $\eta=\kappa_{ext}/\kappa$ is the coupling parameter; $\alpha_f$ is the fano parameter; $g= Gx_{zpf} \sqrt{n_p}$ is the enhanced optomechanical coupling; and $n_p$ is the pump strength expressed in photon units:
\begin{equation}
  n_p = \frac{4P_{in}\kappa_{ext}}{\hbar\omega_p\left( \kappa^2+4\left( \omega_p-\omega_c\right)^2 \right) }
  \label{npLaw}
\end{equation}
 where $P_{in}$ is the incident microwave power.
 
\begin{figure*}
\includegraphics[scale=1]{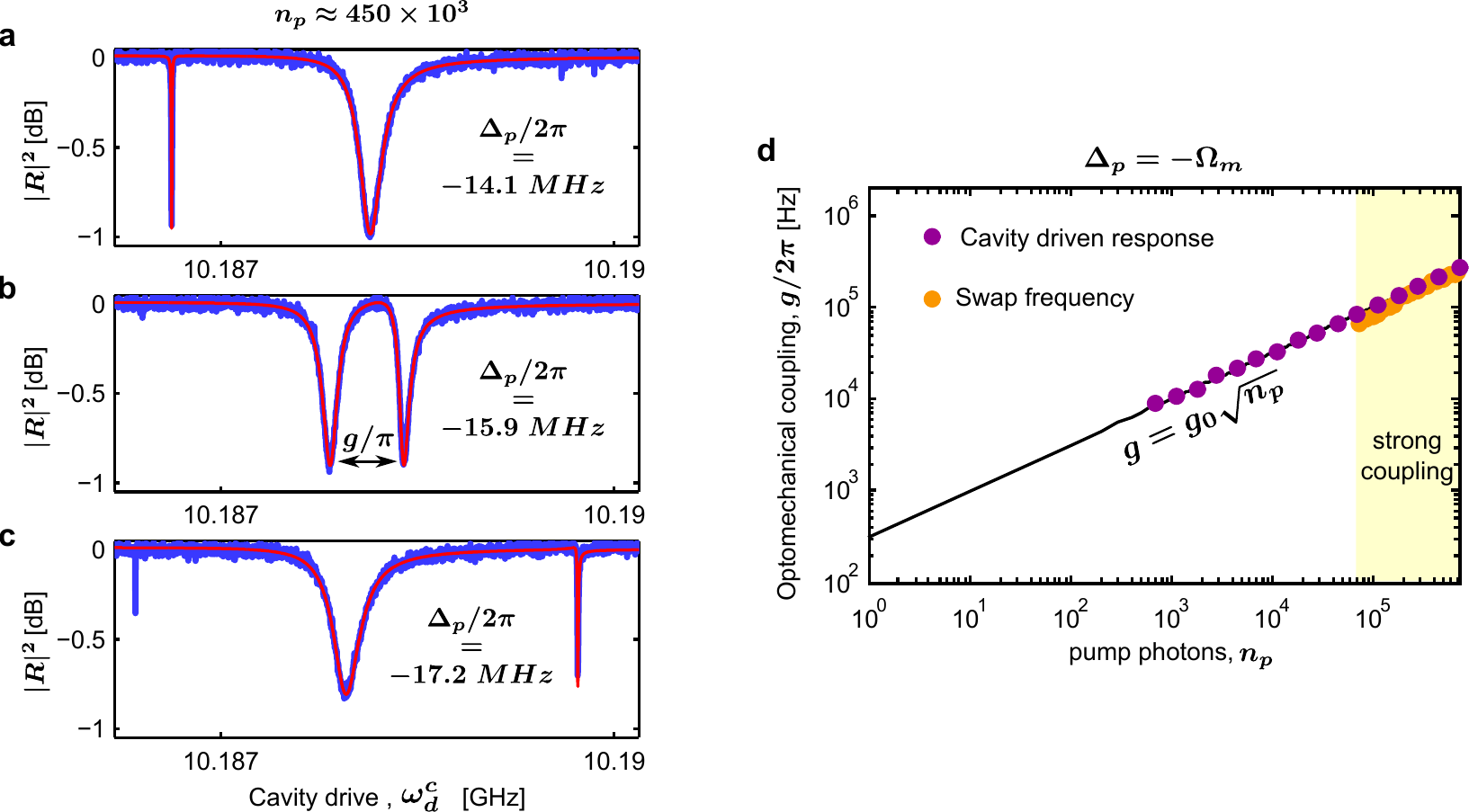}
\caption{\textbf{Cavity driven response and optomechanical coupling}. \textbf{a},\textbf{b},\textbf{c}, Driven response measured in presence of a strong pump for three different detuning and a fixed pump power so that $n_p=447\times10^3$. The fit to Eq.\ref{CavityS11} is shown in red. \textbf{d}, coupling rate as a function of $n_p$. The purple point are measured from the cavity driven response when $\Delta_p=-\Omega_m$, like in \textbf{b}. The orange points are measured from the swap rate between the mechanical oscillator and the microwave cavity, using the same sequence than in Fig.3 and Fig.\ref{OptoMechaVsThetaVsPower}.
\label{OMfreqDomain}}
\end{figure*}

Three individual spectra from Fig.1F and the associated fits to Eq.\ref{CavityS11} are presented in Fig.\ref{OMfreqDomain}, for a fixed pump strength $n_p=447\times10^3$, showing very good agreement between theory and experiment.
 
\begin{figure}
\includegraphics[scale=1]{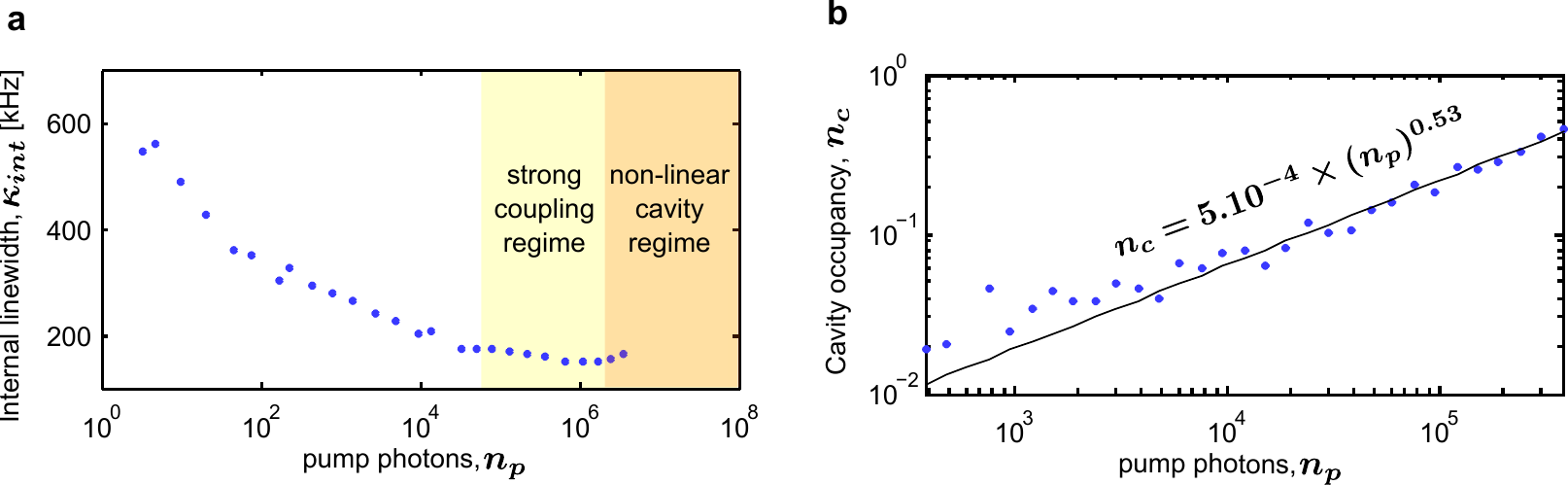}
\caption{\textbf{Signatures of dielectric loss in the microwave cavity}. \textbf{a}, Internal cavity loss as a function of the probe tone strength, expressed in units of intra-cavity photons (see Eq.\ref{npLaw}). \textbf{b}, Excess cavity occupancy measured after a $10~\rm{\mu s}$ pulse at $\Delta_p=-\Omega_m$ with the mechanical mode pre-cooled to $\bar{n}_m^{i}<1$, as a function of pump power. It corresponds to the population measured at the highest interaction angle in Fig.\ref{OptoMechaVsThetaVsPower}e, f, g and h.
\label{HeatingAndKappaVsPower}}
\end{figure}

\begin{figure*}
\includegraphics[scale=0.925]{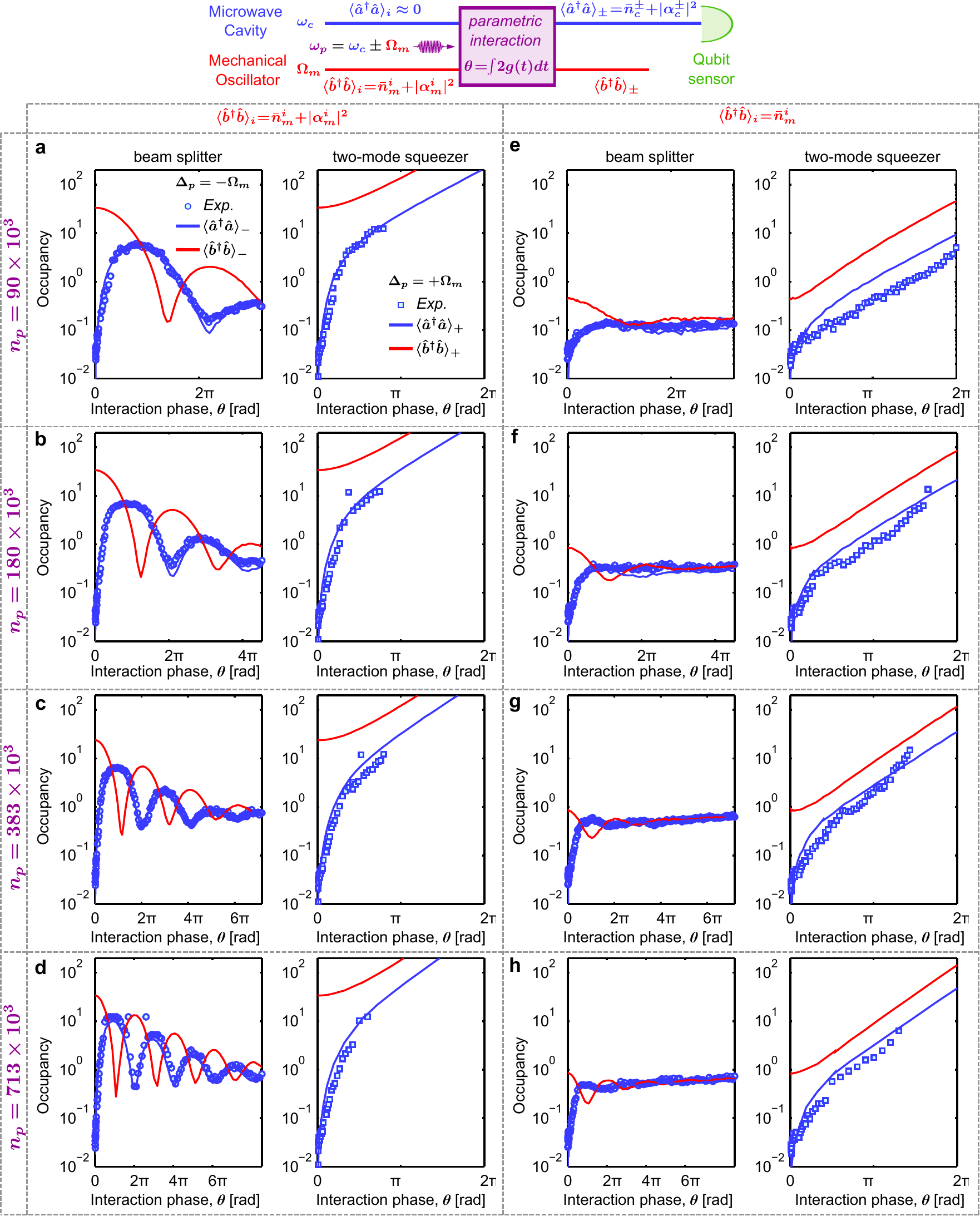}
\caption{\textbf{Pulsed optomechanics with a qubit readout}. Cavity occupancy, in blue, as a function of the interaction duration in reduced units $\theta$, for $\Delta_p=-\Omega_m$ and $\Delta_p=+\Omega_m$, for various pump strengh $n_p$ and initial mechanical state  $\moy{\hat{b}^{\dagger}\hat{b}}_{i}$. Predictions from Eq.\ref{EoM} are shown as solid blue lines. The expected evolution of the mechanical occupancy is shown in red. From \textbf{a} to \textbf{d} and from \textbf{e} to \textbf{h} the pump amplitude increases from $n_p=90\times10^3$ to $n_p=713\times10^3$. In all cases the mechanical mode is initially precooled down to $\bar{n}_m^{i}<1$ and in \textbf{a, b, c, d} it is driven up to $|\alpha_m|^2\approx23$.
\label{OptoMechaVsThetaVsPower}}
\end{figure*}

A fit to the cavity driven response for $\Delta_p=-\Omega_m$ as a function of power allows us to extract the optomechanical coupling  $g$ as a function of the pump strength $n_p$, in purple in Fig.\ref{OMfreqDomain}d. Our data follow the expected $g= g_0 \sqrt{n_p}$ behavior and we extract $g_0/2\pi = 300~\rm{Hz}$. Also, in orange, we plot the optomechanical coupling strength obtained from the swap rate between the microwave cavity and the mechanical oscillator, measured following the same sequence as in Fig.3 and Fig.\ref{OptoMechaVsThetaVsPower}.
We also observe that the internal loss in the cavity improves with the number of intra-cavity photons, as shown in Fig.\ref{HeatingAndKappaVsPower}a. It is a usual behavior in superconducting resonators, and is due to spurious two level systems in the surface dielectrics\cite{Gao2007,Gao2011}. During the strong optomechanical interaction pulses, the cavity lifetime is much longer than at the single photon level and $1/\kappa\approx1~\rm{\mu s}\gg T_{1,c}$.

\textbf{Optomechanics in the time domain}
In this textbf we introduce the theory that describes the out-of-equilibrium dynamics of the optomechanical system. Similar derivations can be found in other references \cite{Clerk2010,Teufel2011b,Palomaki2013a,Aspelmeyer2013,Harlow2013}.
The main goal here is to write the Equations of Motion (EoM) of the system. We will then numerically integrate these EoM and compare the predictions to our data. In the following, the qubit is far detuned from the cavity (up to $\Delta_{qb}\approx2\pi\times3.5~\rm{GHz}\approx280\times J$) and we can neglect its interaction. 

We start with the Hamiltonian describing the coupled cavity/oscillator system:

\begin{equation}
\hat{\mathcal{H}}_{om}= \hbar\omega_c\hat{a}^{\dagger}\hat{a} + \hbar\Omega_{m}\hat{b}^{\dagger}\hat{b} + \hat{\mathcal{H}}_{rp}
\label{OM_hamiltonian}
\end{equation}

where $\hat{\mathcal{H}}_{rp}$ describes the work done by the microwave radiation force onto the mechanical motion:

\begin{equation}
\hat{\mathcal{H}}_{rp}= \hbar g_0\hat{a}^{\dagger}\hat{a}\left(\hat{b}^{\dagger}+\hat{b}\right)
\end{equation}

The Heisenberg-Langevin EoM using input-output formalism are \cite{Clerk2010}:

\begin{flalign}
&\dot{\hat{a}} (t)=-\frac{i}{\hbar}\left[\hat{a},\hat{\mathcal{H}}_{om}\right] - \frac{\kappa}{2}\hat{a} + \sqrt{\kappa_{ext}}\hat{\xi}_{ext} + \sqrt{\kappa_{int}}\hat{\xi}_{int} \label{EoM_1}\\
&\dot{\hat{b}} (t)=-\frac{i}{\hbar}\left[\hat{b},\hat{\mathcal{H}}_{om}\right] - \frac{\Gamma_{m}}{2}\hat{b} + \sqrt{\Gamma_{m}}\hat{\xi}_{m}  \label{EoM_2}
\end{flalign}

which includes the input noise operators associated with the different baths of the system: the external control line of the cavity $\hat{\xi}_{ext}$, the internal bath of the cavity $\hat{\xi}_{int}$ and the internal bath of the oscillator $\hat{\xi}_{m}$.

To capture the dynamics of the coupled modes amplitudes in presence of a strong classical microwave pump, we proceed to the transformation:

\begin{flalign}
&\hat{a} \rightarrow \alpha e^{i\omega_pt} + \hat{a}e^{i\omega_ct}\\
&\hat{b} \rightarrow b_0 + \hat{b}e^{i\Omega_mt}
\end{flalign}

where $\omega_{p}$ and $\alpha$ are the frequency and amplitude of the pump; $b_0$ is the steady state mechanical position in presence of the pump and $\hat{a}$ and $\hat{b}$ describe now the fluctuation around the steady state at the cavity frequency and mechanical frequency, respectively. Note that $\omega_c$ is renormalized to $\omega_c \rightarrow \omega_c + g_0(b_0+b_0^*)$, without a change of notation, for clarity.  We will assume $\alpha$ is real and $\alpha=|\sqrt{n_p}|\gg1$. We can now linearize the Hamiltonian $\hat{\mathcal{H}}_{rp}$ by expanding in powers of $\alpha$, keeping the terms of orders $|\alpha|^1$. Although relatively complex in general the interaction Hamiltonian can take very simple forms for the following specific cases.

Case 1: when $\Delta_p=\omega_p-\omega_c=-\Omega_m$, we get under the rotating wave approximation
\begin{equation} 
\hat{\mathcal{H}}_{rp}\approx\hat{\mathcal{H}}_{-}=\hbar g\left( \hat{a} \hat{b}^{\dagger} + \hat{b} \hat{a}^{\dagger}\right).
\end{equation}

Case 2: when  $\Delta_p=\omega_p-\omega_c=+\Omega_m$, we get under the rotating wave approximation
\begin{equation} 
\hat{\mathcal{H}}_{rp}\approx\hat{\mathcal{H}}_{+}=\hbar g\left( \hat{a}^{\dagger} \hat{b}^{\dagger} +  \hat{b} \hat{a} \right).
\end{equation}

Here $g=g_0\alpha$ is the parametrically enhanced optomechanical coupling.
These simplifications of $\hat{\mathcal{H}}_{rp}$ can be understood from a simple energy conservation argument. When $\omega_p=\omega_c+\Omega_m$ only the parametric down conversion of pump photons into cavity photons and mechanical phonons conserves energy ($\hat{\mathcal{H}}_{+}$). When  $\omega_p=\omega_c-\Omega_m$ the pump mediates the exchange interaction, also known as beam splitter $\hat{\mathcal{H}}_{-}$, by allowing the combination of mechanical phonons and pump photons to generate cavity photons.

To capture the dynamics in the presence of pulsed optomechanical interactions, we now consider a time-dependent number of pump photons $n_-(t)$ and $n_+(t)$, respectively at the pump frequencies $\Delta_p=-\Omega_m$ and $\Delta_p=+\Omega_m$, leading to two different coupling rates $g_{\pm}(t)=g_0|\sqrt{n_{\pm}(t)}|$. The coupled EoM, Eq.\ref{EoM_1} and Eq.\ref{EoM_2}, becomes:

\begin{widetext}
\begin{equation}
\frac{d}{dt}  
  \left[ \begin{array}{c}    
  \hat{a} \\ \hat{b} \\ \hat{a}^{\dagger} \\ \hat{b}^{\dagger}
  \end{array} \right]  
  =
  \left[ \begin{array}{cccc}        
  -\kappa/2 & ig_{-}(t) & 0 & ig_{+}(t)\\
  ig_{-}(t) & -\Gamma_m/2 & ig_{+}(t) & 0\\
  0 & -ig_{+}(t) & -\kappa/2 & -ig_{-}(t)\\
  -ig_{+}(t) & 0 & -ig_{-}(t) & -\Gamma_m/2    
  \end{array} \right]    
  \left[ \begin{array}{c}      
  \hat{a} \\ \hat{b} \\ \hat{a}^{\dagger} \\ \hat{b}^{\dagger}   
  \end{array} \right] 
  +  
  \left[ \begin{array}{cccccc}        
    \sqrt{\kappa_{ext}} & \sqrt{\kappa_{int}} & 0 & 0 & 0 & 0\\
    0 & 0 & \sqrt{\Gamma_m} & 0 & 0 & 0  \\
    0 & 0 & 0 & \sqrt{\kappa_{ext}} & \sqrt{\kappa_{int}} & 0\\
    0 & 0 & 0 & 0 & 0 & \Gamma_m    
    \end{array} \right]  
  \left[ \begin{array}{c}        
  \hat{\xi}_{ext} \\ \hat{\xi}_{int} \\ \hat{\xi}_{m} \\ \hat{\xi}^{\dagger}_{ext} \\ \hat{\xi}^{\dagger}_{int} \\ \hat{\xi}_{m}^{\dagger}     
  \end{array} \right]  
\label{EoM}
\end{equation}
\end{widetext}

The average cavity occupancy is $\moy{\hat{a}^{\dagger}\hat{a}}$ and the average mechanical occupancy is $\moy{\hat{b}^{\dagger}\hat{b}}$. Here we will consider displaced thermal states for both the microwave cavity and mechanical oscillator. The displacements $\alpha_{c,m}$ are given by $|\alpha_{c}|=|\moy{\hat{a}}|$ and $|\alpha_{m}|=|\moy{\hat{b}}|$. The thermal occupancies $\bar{n}_{c,m}$ are given by $\bar{n}_{c}=\moy{\hat{a}^{\dagger}\hat{a}}-|\alpha_c|^2 $ and $\bar{n}_{m}=\moy{\hat{b}^{\dagger}\hat{b}}-|\alpha_m|^2$.

We can now numerically integrate Eq.\ref{EoM} to predict the behavior of $\bar{n}_{c,m}$ and $\alpha_{c,m}$: 
\begin{itemize}
\item The circuit parameters are known from the measurements in the frequency domain described in the previous textbf.
\item The white noise operators satisfy the relations: $\langle \hat{\xi}_{ext,int,m}^{\dagger}(t) \hat{\xi}_{ext,int,m}(0) \rangle = \bar{n}_{ext,int,m}^{eq} \delta(t)$ and $\langle \hat{\xi}_{ext,int,m}(t) \hat{\xi}_{ext,int,m}^{\dagger}(0) \rangle = (\bar{n}_{ext,int,m}^{eq}+1) \delta(t)$. Here: $\bar{n}_{m}^{eq}$ is the equilibrium Bose-Einstein occupancy $\bar{n}_m^{eq}=\left[exp(\hbar\Omega_m/k_bT)-1\right]^{-1}\approx32$ for $T=25~\rm{mK}$;  $\bar{n}_{ext}^{eq}=0$; and, although the environment temperature for the microwave cavity mode, $\bar{n}_{int}^{eq}$,  should be negligible it appears to be finite and dependent on the pump strength, as measured independently in Fig.\ref{HeatingAndKappaVsPower} (discussed further in that textbf).
\item We input the full time domain behavior of $g_{\pm}(t)$, including the Gaussian rise and fall time, as well as a $100~\rm{ns}$ delay corresponding to the time required to bring safely the qubit into resonance with the cavity.
\item The only remaining free parameters are the initial mechanical thermal occupancy and displacement: $\moy{\hat{b}^{\dagger}\hat{b}}_{i}=\bar{n}_{m}^{i} + |\alpha_{m}^i|^2$.
\item We repeat the integration of Eq.\ref{EoM} for $10^4$ initial input states configurations, with a Gaussian distribution of known variance given by initial thermal occupancies and displacements, and extract the moments of the output distribution to estimate $\bar{n}_{c,m}$ and $\alpha_{c,m}$.
\end{itemize} 

In Fig.\ref{OptoMechaVsThetaVsPower} we compare the predictions of the EoM with our data, for four different pump powers and two initial mechanical states, as a function of the interaction duration $\tau_{p}$, expressed in reduced unit $\theta=\int 2g_{\pm}(t)dt$. Note that Fig.\ref{OptoMechaVsThetaVsPower}c is identical to Fig.3b-c in the main text.

\medskip

\textit{Mechanical state preparation:} The rate at which the mechanical oscillator reaches equilibrium with its environment ($\Gamma_{m}/2\pi= 150~\rm{Hz}$) is much slower than our measurement repetition rate of $5~\rm{kHz}$ (see the textbf Experimental setup). That motivates the use of an active reset of the mechanical state, which we will describe now: First we apply a strong microwave pump at $\Delta_p=-\Omega_m$ for $\tau_p\approx20~\rm{\mu s}$ which cools down the mechanical oscillator close to the ground state $\bar{n}_{m}^{i}<1$. For optimized pre-cooling parameters we obtain $\bar{n}_{m}^{i}\approx0.25$ (see Fig.3 and Fig.\ref{OptoMechaVsThetaVsPower}c). We then eventually displace that cold thermal state by applying a second pump at $\Delta_p=-2\Omega_m$ which beats with the cooling pump at the mechanical frequency (see Fig.3 and Fig.\ref{OptoMechaVsThetaVsPower}a-d) and drives the oscillator.

\medskip

\textit{Cavity heating:} One of the recurrent features we observe when measuring the cavity population evolution for $\Delta_p=-\Omega_m$ is that even though the microwave cavity starts cold (initial occupancy $\moy{\hat{a}^{\dagger}\hat{a}}_{i}\approx10^{-2}$) it is driven by excess noise during the interaction, forcing the two modes to equilibrate at a much higher occupancy. The excess noise increases with the pump power, as shown in Fig.\ref{HeatingAndKappaVsPower}a where we measure the final equilibrium occupancy of the cavity $\bar{n}_c=\bar{n}_{int}^{eq}$ for various pump power. At the highest pump powers the occupancy of the cavity reaches a few tenths of quanta.
To explain this excess noise we start by ruling out simple origins like phase or amplitude noise of the pump (linear with power and already filtered out, as shown in Fig.\ref{CavityFilter}). To rule out Joule heating at the feedline port's load we note that the cavity is strongly under-coupled, $\eta=\kappa_{ext}/\kappa_{int}<10\%$, which implies that the feedline port would have to have a temperature of $T_{ext} = T_{cav}/\eta>4~\rm{K}$, unrealistic experimentally. We also rule out Joule heating of the substrate since the qubit would also be affected, but this is not observed. Finally, we converge toward a hot bath of spurious TLSs in the surface dielectrics\cite{Gao2007,Gao2011}, consistent with the observed power law in Fig.\ref{HeatingAndKappaVsPower}b and the power dependent internal quality factor of the cavity shown in Fig.\ref{HeatingAndKappaVsPower}a.

\textbf{Measurement of the vacuum fluctuations}
Even though Eq.\ref{EoM} describes very accurately the behavior of the optomechanical system, it is solved numerically and therefore lacks of physical intuition. We describe in this textbf a simpler analytical model, more intuitive, and discuss the advantages of a number distribution measurement to resolve the vacuum fluctuations of the mechanical oscillator's motion and the microwave field, shown in Fig.3d.

Starting from an ideal, lossless case: $\kappa_{int}=\kappa_{ext}=\Gamma_{m}=0$. Eq.\ref{EoM} becomes simply:

\begin{equation}
\frac{d}{dt}  
  \left[ \begin{array}{c}    
  \hat{a} \\ \hat{b} \\ \hat{a}^{\dagger} \\ \hat{b}^{\dagger}
  \end{array} \right]  
  =
  \left[ \begin{array}{cccc}        
  0 & -ig_{-} & 0 & -ig_{+}\\
  -ig_{-} & 0 & -ig_{+} & 0\\
  0 & ig_{+} & 0 & ig_{-}\\
  ig_{+} & 0 & ig_{-} & 0    
  \end{array} \right]    
  \left[ \begin{array}{c}      
  \hat{a} \\ \hat{b} \\ \hat{a}^{\dagger} \\ \hat{b}^{\dagger}   
  \end{array} \right]  
\label{EoMforSB}
\end{equation}

For a beam splitter ($\Delta_p =-\Omega_m$ and $g_+=0$), and defining $\theta=2g_-t$, one gets:
\begin{equation}
\Delta_p =-\Omega_m\Rightarrow g_+=0\Rightarrow\left\{
\begin{array}{rr}
\hat{a}(t) = \hat{a}(0)\cos(\theta/2) \ + \hat{b}(0) \sin(\theta/2) \\
\hat{b}(t) = \hat{b}(0)\cos(\theta/2) \ + \hat{a}(0) \sin(\theta/2)
\end{array} \right.
\label{NvsTime_red}
\end{equation}

For the two-mode squeezer ($\Delta_p =+\Omega_m$ and $g_-=0$), and defining $\theta=2g_+t$, one gets:
\begin{equation}
\Delta_p =+\Omega_m\Rightarrow g_-=0\Rightarrow\left\{
\begin{array}{rr}
\hat{a}(t) = \hat{a}(0)\cosh(\theta/2) \ + \hat{b}^{\dagger}(0) \sinh(\theta/2) \\
\hat{b}(t) = \hat{b}(0)\cosh(\theta/2) \ + \hat{a}^{\dagger}(0) \sinh(\theta/2)
\end{array} \right.
\label{NvsTime_blue}
\end{equation}

 We define the average photon occupancy after a beam splitter interaction, $\moy{\hat{a}^{\dagger}\hat{a}}_{-}$, the average photon occupancy after a two-mode squeezer interaction,$\moy{\hat{a}^{\dagger}\hat{a}}_{+}$, and the initial mechanical and cavity  occupancies, $\moy{\hat{b}^{\dagger}\hat{b}}_{i}$ and $\moy{\hat{a}^{\dagger}\hat{a}}_{i}$. From Eq.\ref{NvsTime_red} and Eq.\ref{NvsTime_blue} one can calculate the evolution of the total mode occupancy:

\begin{flalign}
 & \moy{\hat{a}^{\dagger}\hat{a}}_{-} =\moy{\hat{a}^{\dagger}\hat{a}}_i \,  \cos^2(\theta/2) \ + \moy{\hat{b}^{\dagger}\hat{b}}_i \, \sin^2(\theta/2)\label{NredLossLess} \\
 & \moy{\hat{a}^{\dagger}\hat{a}}_{+} =\moy{\hat{a}^{\dagger}\hat{a}}_i \cosh^2(\theta/2) + \moy{\hat{b}\hat{b}^{\dagger}}_i\sinh^2(\theta/2) \label{NblueLossLess}
\end{flalign}

One can see clearly the periodic behavior of the beam splitter interaction in Eq.\ref{NredLossLess} and the exponential behavior of the two-mode squeezer interaction in Eq.\ref{NblueLossLess}. In addition, in Eq.\ref{NblueLossLess}, one should notice the commutation relation of the mechanical field operators, $\moy{\hat{b}\hat{b}^{\dagger}}_{i}=\moy{\hat{b}^{\dagger}\hat{b}}_{i}+1$. Experimentally, the microwave cavity is originally in the ground state, $\moy{\hat{a}^{\dagger}\hat{a}}_{i}=0$, which implies:
 
\begin{flalign}
& \frac{\moy{\hat{a}^{\dagger}\hat{a}}_{-}}{\sin^2(\theta/2)}=\moy{\hat{b}^{\dagger}\hat{b}}_{i} \label{NredLossLess_gnd} \\
& \frac{\moy{\hat{a}^{\dagger}\hat{a}}_{+}}{\sinh^2(\theta/2)}=\moy{\hat{b}\hat{b}^{\dagger}}_{i}=\moy{\hat{b}^{\dagger}\hat{b}}_{i}+1 \label{NblueLossLess_gnd}
\end{flalign}

If one calculates similarly the final cavity displacement, $|\alpha_{c}^{\pm}|=|\moy{\hat{a}}_{\pm}|$, for both interactions, as a function of the initial mechanical displacement, $|\alpha_{m}^{i}|=|\moy{\hat{b}}_{i}|$, one obtain:

\begin{equation}
\frac{|\alpha_{c}^{-}|^2}{\sin^2(\theta/2)}=\frac{|\alpha_{c}^{+}|^2}{\sinh^2(\theta/2)}=|\alpha_{m}^{i}|^2 \label{Ncoh_LossLess}
\end{equation}

A single added quantum is revealed in Eq.\ref{NblueLossLess_gnd}, arising from the amplification of the mechanical vacuum fluctuations. The gain of each process is ideally ``$\sin^2(\theta/2)$'' for the beam splitter interaction and ``$\sinh^2(\theta/2)$'' for the two-mode squeezer. The absence of commutation relation in Eq.\ref{Ncoh_LossLess} allows us to measure these gains by simply looking at the coherent components.

In order to compare to our data we will now include in this model the finite linewidth of the cavity and mechanical mode. It induces losses and coupling to the different baths:

\begin{itemize}
\item Losses simply reduce the efficiency for each processes. As shown in Eq.\ref{Ncoh_LossLess} this is easily measured independently by preparing a displaced initial mechanical state, and monitoring the cavity displacement. We define $G_{\pm}$ using the \textit{initial} mechanical displacement $|\alpha_m^i|$ and the \textit{final} cavity displacement $|\alpha_c^{\pm}|$, so that $G_{\pm}=|\alpha_c^{\pm}|^2/|\alpha_m^i|^2$, respectively for $\Delta_p =\pm\Omega_m$. Equations \ref{NredLossLess_gnd} and  \ref{NblueLossLess_gnd} become:
\begin{flalign}
& \frac{\moy{\hat{a}^{\dagger}\hat{a}}_{-}}{G_{-}}=\moy{\hat{b}^{\dagger}\hat{b}}_{i} \label{Nred} \\
& \frac{\moy{\hat{a}^{\dagger}\hat{a}}_{+}}{G_{+}}=\moy{\hat{b}^{\dagger}\hat{b}}_{i}+1 \label{Nblue}
\end{flalign}

\item The finite mechanical bath temperature can be neglected because the timescales associated with the mechanical oscillator are small to first order, $\Gamma_m,\Gamma_{th}\ll g,\kappa,\kappa_{th}$, where $\kappa_{th}=\bar{n}_{int}^{eq}\kappa$. To account for the finite cavity bath temperature discussed in the previous textbf, we assume that the small excess cavity occupancy growing during the optomechanical interaction adds up linearly to the total cavity occupancy. In other word we assume that the excess cavity occupancy is not being transfered by $\hat{\mathcal{H}}_{-}$, nor amplified by $\hat{\mathcal{H}}_{+}$. This approximation is valid in our case as the optomechanical interaction is faster than the thermal decoherence rate of the cavity, $g/\kappa_{th}\approx3$. For either the beam splitter or the two-mode squeezer the contribution of the optomechanical interaction alone $\bar{n}_c^{\pm}$ is retrieved from the total cavity occupancy $\bar{n}_c$ following:
\begin{equation}
\bar{n}_c = \bar{n}_c^{\pm}+\bar{n}_{int}^{eq}\left(1-e^{-\kappa_{opt}t}\right)
\label{RemoveHeating}
\end{equation}
Where $\bar{n}_{int}^{eq}$ is the bath occupancy extracted in Fig.\ref{HeatingAndKappaVsPower}b, $\kappa_{opt}$ is the effective cavity linewidth during the optomechanical interaction ($\kappa_{opt}\approx\kappa/2$ when $\Delta_p=-\Omega_m$ and $\kappa_{opt}\approx\kappa$ when $\Delta_p=+\Omega_m$).
\end{itemize}

We finally emphasize that this simplified model is in very good agreement with the full numerical integration of Eq.\ref{EoM}.

\begin{figure}
\includegraphics[scale=1]{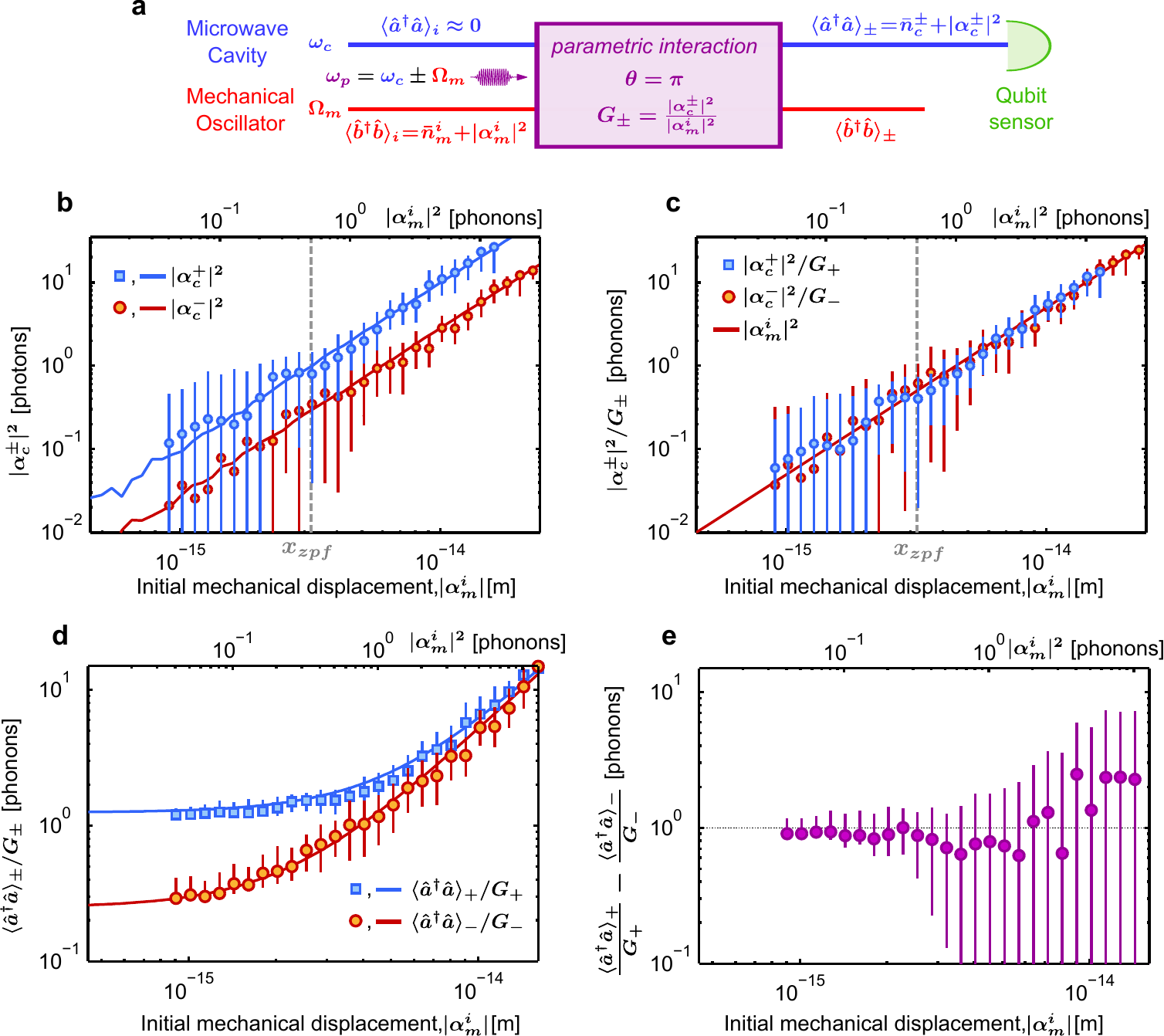}
\caption{\textbf{Measurement of the vacuum fluctuations of the mechanical oscillator.} \textbf{a}, Sequence diagram.\textbf{b}, Measured cavity displacement $|\alpha_{c}^{\pm}|^2$ for each interaction, \textit{i.e} $\Delta_p=\pm\Omega_m$, in red and blue, and corresponding numerical simulations (solid lines).\textbf{c,} Cavity displacement referred back to the input of the optomechanical interaction $|\alpha_{c}^{\pm}|^2/G_{\pm}$, in red and blue, and comparison to the initial mechanical displacement$|\alpha_{c}^{i}|^2$ (red line). \textbf{d}, Total cavity occupancy referred back to the input of the optomechanical interaction $\moy{\hat{a}^{\dagger}\hat{a}}_{\pm}/G_{\pm}$ (spurious cavity heating removed, see text). The solid red and blue lines are respectively the predictions from Eq.\ref{Nred} and Eq.\ref{Nblue}.  \textbf{e,} Difference between the two curves in \textbf{d}, exhibiting the additional quantum due to the amplification of the mechanical fluctuations.
\label{NvsAlpha}}
\end{figure}

We now describe in detail the measurement of Fig.3d (see Fig.\ref{NvsAlpha}). We measure the photon distribution in the cavity (like in Fig.3b-c and Fig.\ref{OptoMechaVsThetaVsPower}) after either a beam splitter or a two-mode squeezer interaction, for a fixed interaction phase $\theta=\pi$ and for a wide range of displacements, $\alpha_m^i$. For negligible displacements we can fit the final cavity distribution to a simple thermal state. As shown in Eq.\ref{RemoveHeating}, we can retrieve the thermal occupancy due the optomechanical interaction alone, $\bar{n}_c^{\pm}$. Then, because the thermal contribution does not change with the mechanical displacement we can then fix that thermal contribution and fit the cavity distribution to a displaced thermal state where the displacement is now the only free parameter. The measured displacements $|\alpha_{c}^{\pm}|^2$ are shown in Fig.\ref{NvsAlpha}b and show good agreement with the numerical simulations (solid lines). In Fig.\ref{NvsAlpha}c we retrieve the initial mechanical displacement $|\alpha_{m}^{i}|^2$ and measure the gains $G_{\pm}$ by plotting $|\alpha_{c}^{\pm}|^2/G_{\pm}$. We can now compare our measurement of $\moy{\hat{a}^{\dagger}\hat{a}}_{\pm}/G_{\pm}$  to Eq.\ref{Nred} and Eq.\ref{Nblue} and observe a good agreement (Fig.3d and Fig.\ref{NvsAlpha}d). In Fig.\ref{NvsAlpha}e we show the difference $\moy{\hat{a}^{\dagger}\hat{a}}_{+}/G_{+}-\moy{\hat{a}^{\dagger}\hat{a}}_{-}/G_{-}$ which exhibit the added quantum originating from the mechanical vacuum fluctuations.

\medskip

\textit{Discussion 1:} A consequence of using a non-linear detector, that measure in the energy basis, is the \textit{direct} access to the occupancy $\langle \hat{a}^{\dagger} \hat{a}\rangle$. To the contrary, when using a linear detector, one measures the quadratures of the light field $(\hat{a}^{\dagger}+\hat{a})/2$ and $i(\hat{a}^{\dagger}-\hat{a})/2$, out of which is inferred $\langle \hat{a}^{\dagger} \hat{a} +  \hat{a}\hat{a}^{\dagger}\rangle = 2\langle \hat{a}^{\dagger} \hat{a}\rangle+1$. In other words, in a number distribution measurement, the vacuum fluctuations are not measured. Although one could recover Eq.\ref{NredLossLess_gnd} Eq.\ref{NblueLossLess_gnd} from the measurement of $\langle \hat{a}^{\dagger} \hat{a} +  \hat{a}\hat{a}^{\dagger}\rangle$ the origin of the additional quantum is very different \cite{Weinstein2014}. From Eq.\ref{NredLossLess} and Eq.\ref{NblueLossLess} one can get:

\begin{flalign}
 & \moy{\hat{a}^{\dagger}\hat{a} + \hat{a}\hat{a}^{\dagger}}_{-} =\moy{\hat{a}^{\dagger}\hat{a}+ \hat{a}\hat{a}^{\dagger}}_i \,  \cos^2(\theta/2) \ + \moy{\hat{b}^{\dagger}\hat{b}+ \hat{b}\hat{b}^{\dagger}}_i \, \sin^2(\theta/2) \label{NredLossLessLin}\\
 & \moy{\hat{a}^{\dagger}\hat{a}+ \hat{a}\hat{a}^{\dagger}}_{+} =\moy{\hat{a}^{\dagger}\hat{a}+ \hat{a}\hat{a}^{\dagger}}_i \cosh^2(\theta/2) +  \moy{\hat{b}^{\dagger}\hat{b}+ \hat{b}\hat{b}^{\dagger}}_i\sinh^2(\theta/2) \label{NblueLossLessLin}  
\end{flalign}  

Notice that now the commutation relations appear symmetrically in both equations. To track the origin of the vacuum fluctuations, let's define $\left[\hat{a},\hat{a}^{\dagger}\right]=\delta_c$ and $\left[\hat{b},\hat{b}^{\dagger}\right]=\delta_m$ where $\delta_c=\delta_m=1$. For a cavity in the ground state, $\moy{\hat{a}^{\dagger}\hat{a}}_{i}=0$, Eq.\ref{NredLossLessLin} and Eq.\ref{NblueLossLessLin} become:

\begin{flalign}
 & \moy{\hat{a}^{\dagger}\hat{a}}_{-} =\moy{\hat{b}^{\dagger}\hat{b}}_i \, \sin^2(\theta/2) + \left(\delta_c\cos^2(\theta/2) + \delta_m\sin^2(\theta/2)- \delta_c\right)/2=\left(\moy{\hat{b}^{\dagger}\hat{b}}_i + (\delta_m- \delta_c)/2\right)\sin^2(\theta/2) \\
 & \moy{\hat{a}^{\dagger}\hat{a}}_{+} =\moy{\hat{b}^{\dagger}\hat{b}}_i \sinh^2(\theta/2) + \left(\delta_c\cosh^2(\theta/2) + \delta_m\sinh^2(\theta/2)- \delta_c\right)/2 =\left(\moy{\hat{b}^{\dagger}\hat{b}}_i + (\delta_m+ \delta_c)/2\right)\sinh^2(\theta/2)
\end{flalign}  

This implies:

\begin{equation}
\frac{\moy{\hat{a}^{\dagger}\hat{a}}_{+}}{\sinh^2(\theta/2)}-\frac{\moy{\hat{a}^{\dagger}\hat{a}}_{-}}{\sin^2(\theta/2)}=\delta_c=1 \label{AsymLinear}
\end{equation}

The asymmetry observed between the two optomechanical interactions is now originating from the vacuum fluctuations of the microwave field and not from the mechanical motion. In fact one would obtain the same asymmetry for a classical mechanical oscillator where $\left[\hat{b},\hat{b}^{\dagger}\right]=\delta_m=0$.

Note that this argument is totally independent of whether the measurement is destructive or "Quantum Non Demolition", and only relies on the fact that we measure energy and not displacement or momentum.

\medskip

\textit{Discussion 2:} In our architecture the qubit measures the intra-cavity field. That has important consequences compared to monitoring the itinerant field exiting the cavity. Indeed, in the latter, correlations between the electromagnetic field noise and mechanical noise can result in "squashing" effects of the output field, complicating the interpretation of the measurements \cite{Jayich2012,Safavi-Naeini2013,Weinstein2014}. Our measurement is not sensitive to that effect.

\begin{figure}  
\begin{minipage}{0.5\linewidth}
\includegraphics[scale=1]{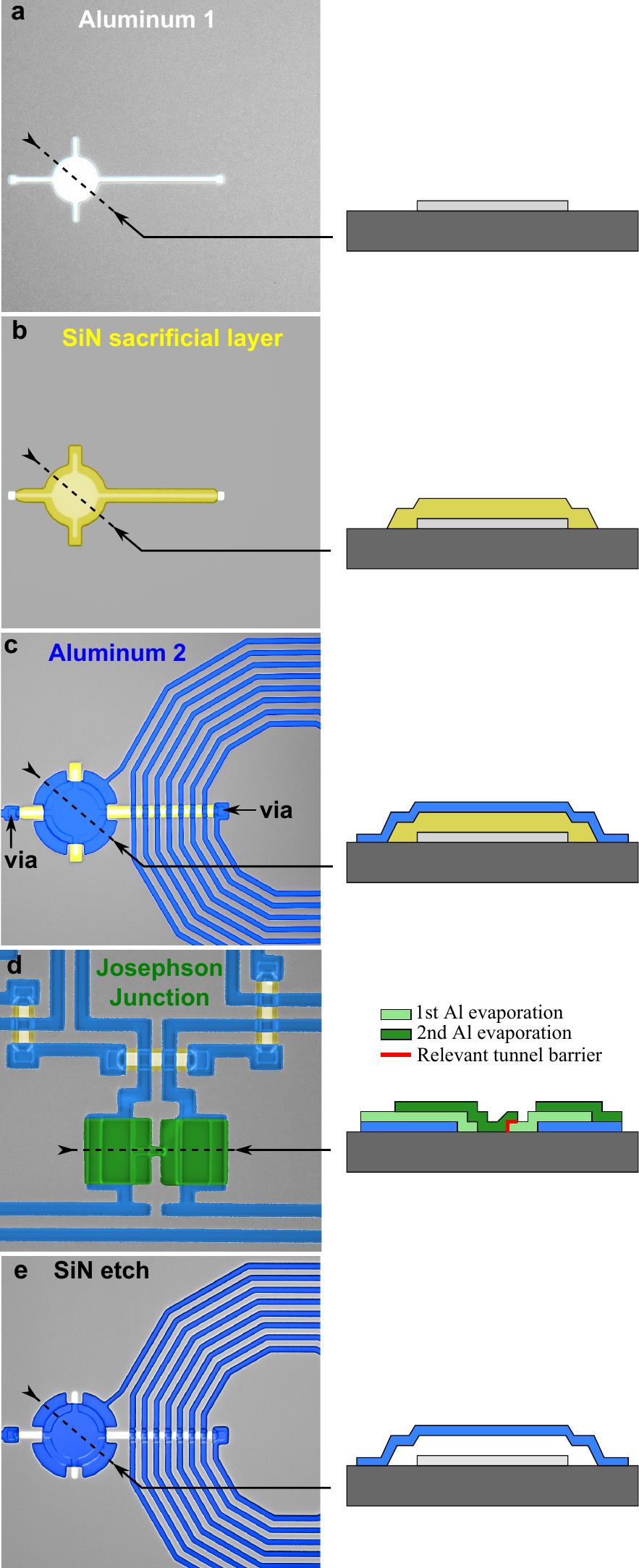}
\end{minipage}
\hspace{0.05\linewidth}
\begin{minipage}{0.40\linewidth}
\caption{\textbf{Device fabrication}. False-colored optical micrograph and side-view diagram at each step of the fabrication process. \textbf{a}, We start by sputtering a $100~\rm{nm}$ thick aluminum layer that is patterned by lithography and etched in a $Cl_2/bCl_3$ plasma. This first layer forms the bottom plate of the cavity capacitor and the central wire of the cavity coil inductor, shown here in white. Also formed in this step are every short interconnection wires in the qubit and SQUID gradiometers (visible in \textbf{d}). \textbf{b}, Next a sacrificial layer is deposited which will give rise to the vacuum gaps in the parallel plate capacitor and other crossovers, shown in yellow. This layer is a $200~\rm{nm}$ thick $SiN_x$ layer that is deposited by PECVD, patterned by lithography and etched in a $CF_4/O_2$ plasma.  \textbf{c}, We sputter and pattern a second aluminum layer ($100~\rm{nm}$ thick, shown in blue, to define the top plate of the capacitor, the cavity coil, qubit coil and capacitor, and almost all the wiring. The electrical contact with the first aluminum layer, in the region not covered with $SiN_x$, is ensured by an in-situ argon plasma etch that removes the aluminum native oxide. \textbf{d}, The Josephson junctions of the SQUID and the qubit (in green) are formed by a double angle aluminum evaporation separated by an in-situ oxidation, using the usual suspended resist mask technique \cite{DOLAN1977}. Again, to ensure proper electrical contact with the previous aluminum layer we ion-mill the native oxide prior to the junction deposition.  \textbf{e}, Finally we release the mechanical membrane and the air bridges by etching the sacrificial layer away, in a high pressure, low power, $SF_6$ plasma.
\label{FabProcess}}
\end{minipage}
\end{figure}

\textbf{Device fabrication}
The device was fabricated on a sapphire substrate using standard optical lithography techniques. The process is illustrated and outlined in 
Fig.\ref{FabProcess}. 

\begin{figure*}
\includegraphics[scale=0.95]{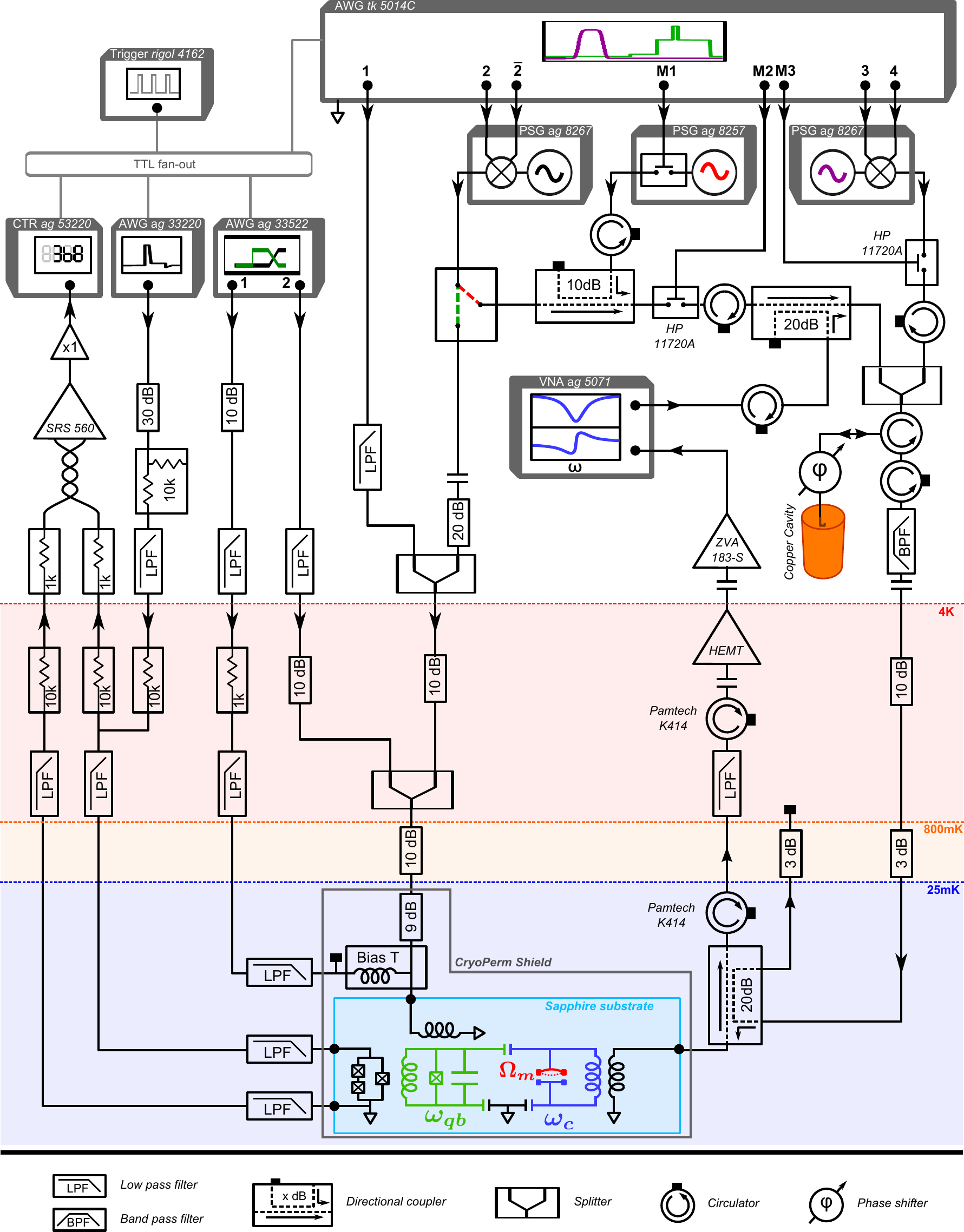}
\caption{\textbf{Detailed experimental diagram}
\label{ExpeDiagram}}
\end{figure*}

\begin{figure}
\begin{minipage}{0.5\linewidth}
\includegraphics[scale=1]{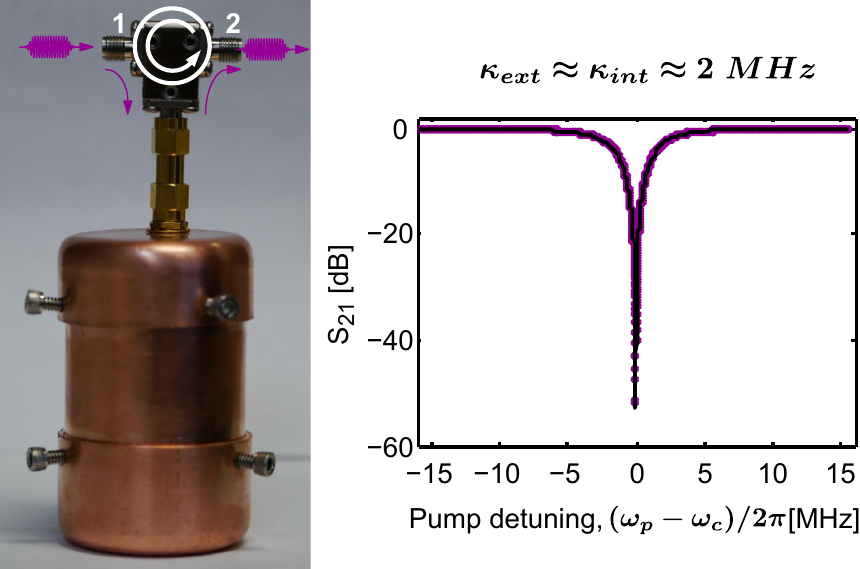}
\end{minipage}
\hspace{0.05\linewidth}
\begin{minipage}{0.40\linewidth}
\caption{\textbf{Copper cavity filter}. Left: picture of the cavity filter, connected to a circulator. Right: Transmitted power through the filter, tuned to resonate at the cavity frequency $\omega_{c}$.
\label{CavityFilter}}
\end{minipage}
\end{figure}

\textbf{Experimental setup}
The chip is wire-bonded to a microwave circuit board, encased in a brass sample box and then anchored to the mixing chamber stage of a dilution refrigerator ($T\approx25~\rm{mK}$).
To protect the circuit from stray magnetic fields we shield the sample box with two concentric Cryoperm cylinders. The dewar itself is also protected by a mu-metal shield.
All the signals are routed down to the device, from room temperature, via coaxial cables. We will describe in the following the signal generation and measurement for the different components of the device. The full experimental setup is shown in 
Fig.\ref{ExpeDiagram}. 
\\

\paragraph*{Microwave optomechanics}
For the control and measurement of the optomechanical device, a total of four different microwave tones are sent down the fridge via one single coaxial line, attenuated at each temperature stage.
The microwave pump (used to produce the parametric coupling between the cavity and the mechanical oscillator) creates the most stringent requirements. The pump strength required to enter the strong optomechanical coupling regime correspond to a large number of intra-cavity photon $n_p>10^5$ at $\Delta_p =\pm\Omega_m$. That pump is generated and pulsed at room temperature using a high power vector signal generator with integrated IQ mixers driven by an Arbitrary Waveform Generator (AWG). This allows for a maximum output power of $25~\rm{dBm}$ \textit{after the mixer}. Here is a list of the requirements:

\begin{itemize}
\item To maintain a negligible intra-cavity population when the pump is off, $n_p\ll10^{-2}$, we need an on/off ratio much bigger than $80~\rm{dB}$. For that we combine the on/off ratio of the generator itself (isolation $>60~\rm{dB}$) with a pin-diode pulse modulator gated by a square pulse (isolation $>60~\rm{dB}$).
\item Because the strong pump is usually applied a mechanical frequency away from the cavity, \textit{i.e} only $90$ cavity linewidth away from resonance, a lot of care was given to avoid non-negligible population at the cavity frequency. First the pump's amplitude and phase noise around $\omega_c$ is filtered out using a home-made notch filter. For that we reflect the pump tone off of a critically coupled copper cavity (see 
Fig.\ref{CavityFilter}). The filter is made from commercial copper tubing, has a quality factor of $\kappa_{int}\approx\kappa_{ext}\approx2~\rm{MHz}$, and the attenuation of the notch filter exceeds easily $50~\rm{dB}$. Second, by pulsing the microwave pump one broadens it in the frequency domain. Simply speaking, if the pump is turned \textit{on} faster than the detuning with respect to the cavity $\Delta_p =\pm\Omega_m$, the cavity gets directly driven, producing unwanted population. To avoid this problem we chose a simple pulse shape with a Gaussian envelope given by the characteristic time $\sigma=200~\rm{ns}$.
\item From Eq.\ref{npLaw}, the pump power in the cavity feedline is close to $0.5~\rm{\mu W}=-33\rm{dBm}$. That high absolute power has two main consequences. First it cannot be dissipated at the mixing chamber stage: a $-13\rm{dBm}$ tone going through a $20~\rm{dB}$ attenuator would dissipate almost $50~\rm{\mu W}$, to be compared to the cooling power of the dilution unit, of the order of a few $\rm{\mu W}$ at base temperature. To avoid that issue we use a $20~\rm{dB}$ directional coupler and route the transmitted signal back to the $T=4~\rm{K}$ stage to be dissipated in a $50~\rm{\Omega}$ load. Second, despite the high output power of the microwave generator ($25~\rm{dBm}$), the attenuation of the coaxial line (from explicit attenuators and cable loss) had to be reduced to its bare minimum without any measurable thermal population of the cavity arising from $300~\rm{K}$ radiation.
\end{itemize}

We combine that main pump tone with two additional microwave drives, used to respectively pre-cool and displace the mechanical mode. The pre-cooling pulse is generated in a very similar way as the main pump and has a fixed frequency, $\omega_{precool}$, so that $\Delta_{precool} =\omega_{precool}-\omega_{c}=-\Omega_m$. To displace the mechanical mode we add another microwave pulse, $\omega_{drive}$, with $\Delta_{drive} =\omega_{drive}-\omega_{c}=-2*\Omega_m$, that beats with the pre-cooling tone at the mechanical frequency, thus directly driving the mechanical mode.

Finally to measure the driven response of the cavity (Fig.2F) we use a Vector Network Analyser (VNA). A weak probe tone is added at room temperature, reflected off of the optomechanical device and amplified back up to room temperature. For these type of measurements the cavity filter is bypassed.
\\

\paragraph*{Qubit control}
The qubit is flux biased via an on-chip coil (mutual inductance $\approx2~\rm{pH}$). The current through the coil comes from three distinct sources. The goal is to access a large range of flux (thus qubit frequency) with a large enough bandwidth, negligible dephasing and no excess dissipation at the mixing chamber stage of the dilution fridge.
\begin{itemize}
\item A very low frequency signal (Bandwidth$\approx2~\rm{MHz}$) is generated at room-temperature with an AWG, then is attenuated and filtered at room temperature, before going through a $1~\rm{k\Omega}$ bias resistor at $T=4~\rm{K}$. The signal is filtered, using commercial low-pass filters filled with lossy copper epoxy, at $T=4~\rm{K}$ (\textit{VLFX-80}) and at base temperature  (\textit{VLFX-80}+\textit{VLFX-650}). That very low frequency signal ensures a stable and non dissipative bias at $\Phi=\Phi_0/2$ during the dcSQUID switching measurement and a bias at $\Phi=0$ otherwise.
\item The two other signals (Bandwidth$\approx30~\rm{MHz}$ and $\approx200~\rm{MHz}$) are generated at room temperature using AWGs, combined at $T=4~\rm{K}$ and routed down via unfiltered, attenuated, coaxial lines. The $\approx30~\rm{MHz}$ bias is used to move from $\Phi=0$, where $\omega_{qb}/2\pi \approx13.5~\rm{GHz}$, to a flux bias where $\omega_{qb}/2\pi \approx11~\rm{GHz}$, about $800~\rm{MHz}$ away from the cavity resonance. The high frequency bias is necessary for tuning the qubit in and out of resonance with the cavity faster than the coupling $J=12.5~\rm{MHz}$. It is also necessary for the qubit tunneling measurement \cite{Cooper2004}.
\end{itemize}
The three signals are finally combined before the sample using of a custom bias-T that has a band-pass going down to dc on all three ports. For qubit state manipulation, microwave pulses are eventually added at room temperature to the high bandwidth bias signal  using a diplexer (cross over frequency at $7.5~\rm{GHz}$).
\\

\paragraph*{dcSQUID readout}
The dcSQUID is current biased by use of a $10~\rm{k\Omega}$ bias resistor at $T=4~\rm{K}$. The bias signal is generated at room temperature by an AWG, followed by a first step of filtering and attenuation. The signal is filtered again at $T=4~\rm{K}$ and at base temperature (same filters than qubit lines). The voltage that develops across the dcSQUID when it switches propagates through identically filtered lines up to room temperature where it is amplified and filtered by a low noise amplifier (\textit{SRS-650}). The voltage goes through a unity-gain isolation amplifier and finally to a counter that records the number of switching events. A $5~\rm{KHz}$ repetition rate ensures proper relaxation of the quasi-particles generated by the dcSQUID switching.

\bibliography{Bibliographie_HybridOptoMecha_V16_Method}

\newpage


\begin{table*}[b]
\begin{tabular}{|c|c|c|}

\hline 
\multicolumn{3}{|c|}{\textit{Device parameters}} \\
\hline

Phase qubit frequency & $\omega_{qb}$ & $9~\rm{GHz} \le \omega_{qb}/2\pi \le 13.5~\rm{GHz}$ \\ \hline 

Microwave cavity frequency & $\omega_{c}$ & $\omega_{c}/2\pi= 10.188~\rm{GHz}$ \\ \hline 

Mechanical oscillator frequency & $\Omega_{m}$ & $\Omega_{m}/2\pi= 15.9~\rm{MHz}$\\ \hline

Qubit-cavity coupling strength & $J$ & $J/2\pi= 12.5~\rm{MHz}$\\ \hline

Cavity sensitivity & $G$ & $G/2\pi= 95~\rm{MHz/nm}$\\ \hline

Mechanical zero-point motion & $x_{zpf}$ & $x_{zpf}= 3.18~\rm{fm}$\\ \hline

optomechanical coupling strength & $g_0$ & $g_0/2\pi= 300~\rm{Hz}$\\ \hline

Mechanical damping rate & $\Gamma_{m}$ & $\Gamma_{m}/2\pi= 150~\rm{Hz}$ \\ \hline 

Internal cavity damping rate & $\kappa_{int}$ & $150~\rm{kHz} \le \kappa_{int}/2\pi \le 1~\rm{MHz}$ \\ \hline 

External cavity damping rate & $\kappa_{ext}$ & $\kappa_{ext}/2\pi= 11~\rm{kHz}$ \\ \hline

Single photon cavity decay time & $T_1^{cav}$ & $T_1^{cav}=130~\rm{ns}$ \\ \hline  

Single photon qubit decay time at $10.5~\rm{GHz}$  & $T_1^{qb}$ & $T_1^{qb}=170~\rm{ns}$ \\ \hline  

Qubit pure dephasing time at $10.2~\rm{GHz}$ & $T_\phi^{qb}$ & $T_\phi^{qb}=45~\rm{ns}$ \\ \hline 

\multicolumn{3}{c}{} \\

\hline  
\multicolumn{3}{|c|}{\textit{Technical parameters}} \\
\hline

Membrane mass & $m$ & $m=52~\rm{pg}$ \\ \hline 

Membrane radius & $r$ & $r=8.25~\rm{\mu m}$ \\ \hline 

Membrane tension & $\sigma$ & $\sigma=320~\rm{MPa}$ \\ \hline 

Membrane spring constant & $k_s$ & $k_s=530~\rm{N/m}$ \\ \hline 

Capacitor plate separation & $d$ & $d=40~\rm{nm}$ \\ \hline 

Membrane capacitance & $C_m$ & $C_m=35~\rm{fF}$ \\ \hline 

Stray capacitance & $C_s$ & $C_s=13~\rm{fF}$ \\ \hline 

Cavity coil inductance & $L$ & $L=5~\rm{nH}$ \\ \hline 

Qubit junction critical current & $I_c^{qb}$ & $I_c^{qb}=0.584~\rm{\mu A}$ \\ \hline 

Qubit shunt capacitance & $C_{qb}$ & $C_{qb}=250~\rm{fF}$ \\ \hline 

Qubit shunt inductance & $L_{qb}$ & $L=1~\rm{nH}$ \\ \hline 

Qubit/antenna mutual inductance & $M_{antenna/qb}$ & $M_{antenna/qb}=2~\rm{pH}$ \\ \hline 

SQUID junctions critical current & $I_{1,2,3}^{squid}$ & $I_{1,2,3}^{squid}=1.75,2.8~\rm{and}~3.15~\rm{\mu A}$ \\ \hline 

SQUID inductance & $M_{squid/qb}$ & $M=300~\rm{pH}$ \\ \hline 

Qubit/SQUID mutual inductance & $M_{squid/qb}$ & $M=70~\rm{pH}$ \\ \hline 

\multicolumn{3}{c}{} \\

\hline  
\multicolumn{3}{|c|}{\textit{Combinations and other parameters}} \\
\hline

Resolve sideband limit factor & $\Omega_m/\kappa$ & $16 \le \Omega_m/ \kappa \le 88$ \\ \hline

Cavity impedance & $Z$ & $Z=380~\rm{\Omega}$ \\ \hline 

Mechanical quality factor & $Q_m=\Omega_{m}/\Gamma_{m}$ & $Q_m = 100,000$ \\ \hline 

Cavity quality factor & $Q_c=\omega_{c}/\kappa$ & $10,000 \le Q_c\le 60,000$ \\ \hline 

Mechanical equilibrium occupancy (T=25mK) & $ \bar{n}_m^{eq} $ & $\bar{n}_m^{eq}=32$ \\ \hline  

\multicolumn{3}{c}{} \\

\hline  
\multicolumn{3}{|c|}{\textit{Various notations}} \\
\hline

Qubit-cavity detuning &\multicolumn{2}{|c|}{ $\Delta_{qb}= \omega_{qb}-\omega_{c}$ }\\ \hline

Pump-cavity detuning & \multicolumn{2}{|c|}{ $\Delta_{p}= \omega_{p}-\omega_{c}$ }\\ \hline

Cavity occupancy & \multicolumn{2}{|c|}{ $\langle \hat{a}^{\dagger} \hat{a}\rangle$} \\ \hline

Mechanical occupancy & \multicolumn{2}{|c|}{ $\langle \hat{b}^{\dagger} \hat{b}\rangle$} \\ \hline

Cavity displacement & \multicolumn{2}{|c|}{ $|\alpha_c|=|\moy{\hat{a}}|$} \\ \hline

Mechanical displacement & \multicolumn{2}{|c|}{ $|\alpha_m|=|\moy{\hat{b}}|$} \\ \hline

Cavity thermal occupancy & \multicolumn{2}{|c|}{ $\bar{n}_c=\moy{\hat{a}^{\dagger}\hat{a}}-|\alpha_c|^2$} \\ \hline

Mechanical thermal occupancy & \multicolumn{2}{|c|}{ $\bar{n}_m=\moy{\hat{b}^{\dagger}\hat{b}}-|\alpha_m|^2$} \\ \hline

\end{tabular}
\caption{\textbf{Device parameters and notations}
\label{DeviceParam}}
\end{table*}

